\documentclass[twocolumn,preprintnumbers]{revtex4}
\usepackage{amssymb}
\usepackage{amsfonts}
\usepackage{amsmath}
\usepackage{graphicx}
\usepackage{bm}
\usepackage{hyperref}
\usepackage{version}

\input{tcilatex}
\begin{document}

\title{Sequential large momentum transfer exploiting rectangular Raman pulses%
}
\author{B. Dubetsky}
\affiliation{Independent Researcher, 1849 South Ocean Drive, Apt 207, Hallandale, Florida
33009, USA}
\email{bdubetsky@gmail.com}
\date{\today }

\begin{abstract}
It is proposed to use rectangular Raman pulses for the technique of
sequantial large momentum transfer. It is shown that the small parameters
that make it possible to use this technology for precision atom
interferometry can be 40--200 times smaller than in the case of the Bragg
regime. It is predicted that in the case of a non-equidistant timing of
auxiliary pulses, one can observe oscillations in time of the interference
picture with a period inversely proportional to the recoil frequency. Such
an observation would be the first confirmation that Mach-Zehnder atom
interference is a phenomenon caused by the quantization of the atomic
center-of-mass motion. This effect is calculated for any shape of pulses.
One can observe it in the Bragg regime as well. It is proposed to use
non-continuous composite Raman pulses as auxiliary beam splitters so that
the effective Rabi frequency remains unchanged for the entire process. The
gravity phase of an atomic interferometer is calculated for any shape,
duration, and timing of Raman pulses, including the Bragg regime. The phase
corrections caused by the finite pulses' durations are also calculated for
the rectangular Raman pulse shape.
\end{abstract}

\maketitle

\twocolumngrid%

\section{\label{s1}Introduction}

Since its birth about 40 years ago \cite{c1}, the field of atom
interferometry has matured significantly. The current state and prospects in
this area are presented, for example, in the reviews in \cite%
{c1.0,c1.1,c1.1.6,c1.1.7,c1.1.5,c1.1.4,c1.1.3,c1.1.2,c1.1.1,c1.1.0} and the
proposals in \cite%
{c1.2,c1.3,c1.4,c1.5,c1.5.1,c1.5.2,c1.5.3,c1.5.6,c1.5.5,c1.5.4}. For the
successful implementation of these programs, it is important to increase the
phase of the atomic interferometer (AI), without enlarging the error of the
phase measurement. Such an increase can be achieved by using AIs with a long
interrogation time $T$ and a large momentum transfer (LMT) during the
interaction of an atom with pulses of optical fields. The highest value of $%
T=1.15$s was achieved \cite{i2} for freely falling atoms, while for atoms
trapped in an optical lattice, the time between the first and last Raman
pulses was increased to 1 min \cite{i3}.

One may consider the non-linear interaction of atoms with a resonant
standing wave pulse as the first implementation of the LMT technique. In
this case, owing to the LMT, higher spatial harmonics of the atomic density
arise [see Eq. (4) in \cite{c1}]. Various modifications of LMT used to
produce such harmonics are briefly described in the Appendix \ref{A1}.

Using the combination of adiabatic rapid passage and multiphoton Bragg
diffraction allowed one to achieve an LMT of $5\hbar k$ \cite{i3.-1}, where $%
k=\left\vert \mathbf{k}\right\vert ,$ with $\mathbf{k}$ the effective wave
vector associated with the atomic beam splitter. The theory of multiphoton
Bragg diffraction was developed and calculation of the phase of the AI at an
LMT\ of $5\hbar k$ was implemented in the article \cite{i3.-1.1}. It was
shown \cite{i3.-1.1.1} how the Bloch oscillation technique leads to an LMT
beam splitter, and a beam splitter having an LMT of $5\hbar k$ was
experimentally demonstrated. An efficient scheme based on fast adiabatic
passage at an LMT of $10\hbar k$, was proposed in \cite{i3.-1.2}. A Raman
beam splitter, having an LMT of $16\hbar k$ \cite{i8}, allowed\ one to
increase the recoil line splitting \cite{i9} by 15 times. Owing to the Bragg
diffraction many times repeated, a beam splitter having an LMT of $45\hbar k$
\cite{i10} coherently splits the atomic cloud into two components separated
from each other by a distance of $54$ cm. In Ref. \cite{i3.0.25} a
three-path AI with an LMT of $56\hbar k$ was created for precision
measurements of the recoil frequency. Exploiting optical lattices as
waveguides and beam splitters promises an LMT of $100\hbar k$ or more \cite%
{i3.0.50}. Instead of Bragg diffraction, one can use pulses of a traveling
wave, resonant to the transition between the ground and metastable excited
states. This approach was realized in Ref. \cite{i14}, where an LMT of $%
146\hbar k$ was created. Twin-lattice atom interferometry leads to an LMT of 
$204\hbar k$ \cite{i3.0.100}. The combination of the fifth order Bragg
scattering and the pulse of the Bloch oscillations allowed one to increase
the LMT to $405\hbar k$ \cite{i3.0.150}. Using the pulse of the Bloch
oscillation \cite{a}, one achieved an LMT of $500\hbar k$ \cite{i3.0.30}.
Owing to these achievements, currently, LMT is one of the promising methods
for improving the accuracy of atomic navigators \cite{i4,i5}, gravimeters 
\cite{i6}, and gyroscopes \cite{i7}. The results of the gravity acceleration
and rotation rate measurements are summarized in the review \cite{c1.1.7}.

\subsection{Sequential large momentum transfer}

One of the varieties of LMT is the sequential method, which uses a sequence
of $\pi -$pulses having opposite effective wave vectors. Sequential LMT
(SLMT) was studied in Refs. \cite{i8,i10,i14}, and in articles \cite{i10,i14}
SLMT was successfully used for the Mach-Zehnder AI (MZAI). One can use three
types of beam splitters here:

\begin{enumerate}
\item[$I$] when the internal state of the atom stays the same during
interaction with a pulse of a resonant optical field;

\item[$II$] when, during a one-photon transition, an atom is excited from
the ground to a metastable excited state;

\item[$III$] when the Raman pulse transfers the atom from one (ground) to
another (excited) sublevel of the ground state.
\end{enumerate}

In Ref. \cite{c1} the type I beam splitter was a standing wave. Atom
interference in the field of a standing wave was observed in Refs. \cite%
{i10.1,i11}. In the Bragg regime, both a standing wave \cite{i12} and
counter-propagating waves with a specially chosen frequency difference \cite%
{i10} can be used. Either a standing \cite{c1} or a traveling wave \cite{i13}
was used as a type-II beam splitter. Type-II SLMT was observed in Ref. \cite%
{i14}. However, for most of the precision gravimeters and gyroscopes listed
in the review in \cite{c1.1.7}, Raman pulses, a type-III beam splitter, were
used.

One can find a comparison of type-I and -III interferometers in Ref. \cite%
{i14.1}, which considers the case when two atomic beam splitters having
opposite effective wave vectors act simultaneously. This field
configuration, the Raman standing wave, was proposed in Ref. \cite{i45}. It
is now better known as the double-diffraction technique \cite{i46,i46.1}. A
further development of this approach, a combination of three Raman beam
splitters, was proposed in Ref. \cite{i14.2}.

For SLMT, it is important that under the action of a $\pi -$pulse, the
momentum of the atom does not change, or changes only by $\pm \hbar \mathbf{k%
}$. For a type-I splitter, this can only be achieved in the Bragg regime,
when the small parameter of the problem is given by 
\begin{equation}
\varepsilon =\left( \omega _{k}\tau \right) ^{-1}\ll 1,  \label{1}
\end{equation}%
where $\tau $ is the pulse duration and%
\begin{equation}
\omega _{k}=\dfrac{\hbar k^{2}}{2M}  \label{2}
\end{equation}%
is the recoil frequency, with $M$ the mass of the atom. Since there are many 
$\pi -$pulses in the SLMT, no matter how small the parameter \ref{1} is,
corrections to the atomic wave function, repeated many times, can lead to
significant changes in the interference pattern. In the experiment \cite{i10}%
\begin{equation}
\varepsilon \approx 0.18.  \label{3}
\end{equation}%
Unlike Ref. \cite{i46.2}, we do not consider here issues related to
frequency noise, or pulse fidelity. Even with beam splitters that are ideal
in these respects, $\pi -$pulses can be imperfect just because their
duration is not long enough.

\begin{figure}[!t]
\includegraphics[width=8cm]{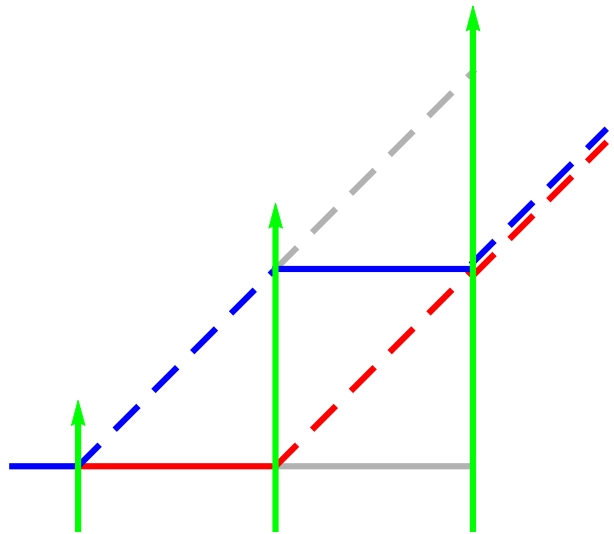}
\caption{Recoil diagram for the MZAI. Atoms in the ground and excited states
are shown by solid and dashed lines, respectively.}
\label{f1}
\end{figure}

The disadvantage of beam splitters III, Raman pulses, are the ac-Stark
shifts of the atomic levels that do not coincide. On the other hand, SLMT
can be realized for any value of $\varepsilon $. Raman pulses can have both
long and short durations corresponding to the Bragg regime $\left(
\varepsilon \ll 1\right) $ and the opposite Raman-Nath regime 
\begin{equation}
\varepsilon \gg 1.  \label{3.1}
\end{equation}%
In this article, we will consider SLMT with Raman pulses only. In this case,
the MZAI scheme is shown in Fig. \ref{f1}.

If the mirror pulse (second Raman pulse) is a resonant $\pi -$pulse, then
the atoms change their momenta and internal states with probability 1. In
this case, the diagram has only two branches, red and blue. Atomic
interference is the interference of the amplitudes of an atom in an excited
state $e$, which arise when an atom moves along these branches. In this
case, the contrast of the interference pattern is equal to $1$. If the
second pulse is not ideal, then the atoms do not change their state with a
small amplitude (see the gray lines in the recoil diagram). Obviously, this
only leads to a decrease in contrast and does not affect the phase of the
MZAI.

\begin{figure}[!t]
\includegraphics[width=8cm]{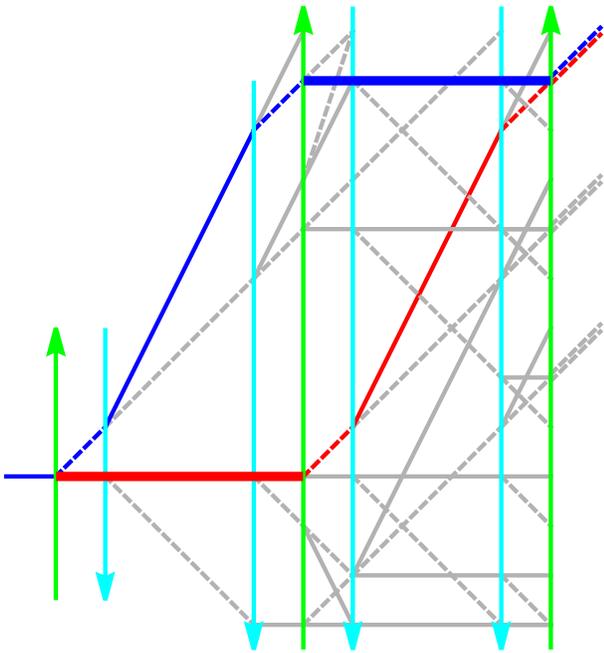}
\caption{SLMT. Each of the four sets of auxiliary pulses contains one $%
\protect\pi -$pulse. The main and auxiliary pulses are shown with green and
cyan arrows, respectively. Thin, blue or red lines correspond to the
resonant branches of the diagram, along which the momentum of the Raman
field $\pm \mathbf{\hbar }\mathbf{k}$ is transferred to the atom and the
internal state of the atom changes. The thick horizontal lines correspond to
the non-resonant branches of the diagram, along which the atoms remain in
the ground state $g$, and the momentum of the atom also does not change.}
\label{f2}
\end{figure}

The situation changes for SLMT, when in addition to the three main Raman
pulses there are four sets of auxiliary $\pi -$pulses, after the first,
before and after the second and before the third main pulses. Consider the
simplest case, when each of the sets consists of only one $\pi -$pulse (see
Fig. \ref{f2}).

Each of the auxiliary pulses must satisfy the two requirements \cite{i10}:

\begin{enumerate}
\item \label{I}It must be a resonant $\pi -$pulse on the resonant branch of
the recoil diagram;

\item \label{II}One has to choose the pulse parameters in such a way that
the state of the unexcited atom on the other (nonresonant) branch remains
unchanged.
\end{enumerate}

If both requirements are met exactly, then, by successively applying
auxiliary pulses $n$ times with alternative wave vectors $\pm \mathbf{k}$,
one obtains a beam splitter with momentum transfer%
\begin{equation}
\Delta \mathbf{p}=\left( n+1\right) \hbar \mathbf{k},  \label{4}
\end{equation}%
and the excitation probability of an atom in a uniform gravity field $%
\mathbf{g}$ oscillates as 
\begin{subequations}
\label{5}
\begin{eqnarray}
w &=&\dfrac{1}{2}\left[ 1-\cos \left( \alpha +\phi _{g}\right) \right] ,
\label{5a} \\
\phi _{g} &=&\left( n+1\right) \mathbf{k}\cdot \mathbf{g}T^{2},  \label{5b}
\end{eqnarray}%
where the parameter $\alpha $, like the parameters $\alpha ^{\prime }$ and $%
\alpha ^{\prime \prime }$ below in Eqs. (\ref{8}, \ref{14}), are parts of
the AI phase, independent of gravity.

In the Raman-Nath regime (\ref{3.1}), when the pulse is so short that the
violation of resonance conditions becomes insignificant, one can ignore both
the first and the second requirements. In this case, the SLMT is twice as
efficient, leading to the momentum transfer 
\end{subequations}
\begin{equation}
\Delta \mathbf{p}=\left( 2n+1\right) \hbar \mathbf{k}.  \label{7}
\end{equation}%
The SLMT in the Raman-Nath approximation was observed in Ref. \cite{i14} for
the one-photon transition. We are not aware of Raman beam splitters
operating in the Raman-Nath regime (\ref{3.1}).

Condition \ref{II} can be satisfied in the Bragg regime (\ref{1}). In this
case, if the Raman pulse is resonant to a transition along one of the
branches, then it is not resonant for the other branch, and with accuracy (%
\ref{1}) one can assume that the state of the atom remains unperturbed along
the other branch. If the pulse is not perfect, and the parameter $%
\varepsilon $ is not small enough, then the pulse ceases to be a mirror,
splits the states of the atom, and parasitic gray trajectories appear (see
Figs. \ref{f1} and \ref{f2}). One sees that, unlike the usual MZAI, in the
MZAI with one auxiliary $\pi -$pulse, parasitic trajectories lead to the
appearance of parasitic interference ports, due to which a small addition to
the signal (\ref{5}) arises, which oscillates like%
\begin{equation}
\Delta w=\beta \cos \left( \alpha ^{\prime }+\mathbf{k}\cdot \mathbf{g}%
T^{2}\right) .  \label{8}
\end{equation}%
It should be emphasized that for precision measurements it is not enough
that this correction be small, it is necessary that the amplitude of this
correction be less than the accuracy of the MZAI phase measurement $\phi _{%
\mathrm{err}}$%
\begin{equation}
\left\vert \beta \right\vert <\phi _{\mathrm{err}}.  \label{9}
\end{equation}%
Only in this case does SLMT lead to progress in improving the accuracy of
precision measurements. The error in the differential measurements of the
MZAIs phases was \cite{i15}%
\begin{equation}
\phi _{\mathrm{err}}\approx 2zc10^{-5}\text{rad.}  \label{10}
\end{equation}%
One sees from Fig. \ref{f2} that parasitic ports are spatially separated
from the main port by a distance%
\begin{equation}
\Delta z=v_{r}T,  \label{11}
\end{equation}%
where $v_{r}=\left\vert \mathbf{v}_{r}\right\vert $ and%
\begin{equation}
\mathbf{v}_{r}=\dfrac{\hbar \mathbf{k}}{M}  \label{12}
\end{equation}%
is the recoil velocity. Under the experimental conditions \cite{i10}, $%
k=1.61\times 10^{7}$ m$^{-1}$, $M=87$ a.u., and $T=1.04$ s, the ports are
separated from each other by a distance of $\Delta z=1.2$ cm. If, despite
thermal expansion, the size of the atomic cloud, as well as the size of the
detector, is less than $\Delta z$, then one can exclude the influence of
parasitic ports.

The situation changes if several $\pi -$pulses are used in each auxiliary
set, $n>1$. The case $n=2$ is shown in Fig. \ref{f3}. One sees that the
distance between the main and parasitic ports decreases to the value

\begin{equation}
\Delta z=v_{r}d,  \label{13}
\end{equation}%
where $d$ is the delay between adjacent auxiliary pulses. At $d=200~\mu $s 
\cite{i16} $\Delta z\approx 2.4~\mu .$ One may encounter technological
difficulties in creating detectors and atomic clouds of such a small size.
Otherwise, the ports overlap and parasitic signals occur. For example, a
port caused by the interference of the blue-orange and red-purple branches
(see Fig. \ref{f3}) results in a parasitic signal 
\begin{equation}
\Delta w=\beta \cos \left( \alpha ^{\prime \prime }+2\mathbf{k}\cdot \mathbf{%
g}T^{2}\right) .  \label{14}
\end{equation}

With a larger number of auxiliary pulses, the role of parasitic terms can
increase even more. Thus, for precision interferometry, it is necessary to
use such beam splitters that satisfy requirements \ref{I} and \ref{II} with
the greatest accuracy. To satisfy requirement \ref{1}, it is enough to
adjust the Raman frequency detuning of a given pulse to the frequency of the
transition between the atomic momentum states before and after the action of
the field, while for requirement \ref{II}, we propose in this paper to use
Rabi oscillations \cite{i17} instead of the Bragg regime \cite{i10}. For the
nonresonant branch, one can find such a pulse duration%
\begin{equation}
\tau \sim \omega _{k}^{-1},  \label{15}
\end{equation}%
at which the probability of state splitting on this branch will be exactly
0. Below we find this duration for a rectangular Raman pulse and calculate
the phase of the MZAI.

Since the pulse on the resonant branch must have an area $\pi $, the
effective Rabi frequency $\Omega $ of a two-quantum transition between
atomic states for auxiliary pulses may not coincide with the Rabi frequency
for the main pulses. This, to a certain extent, is a technological
challenge, the implementation of SLMT with sets of pulses having different
intensities and durations. To circumvent this difficulty, we propose to use
composite pulses \cite{i18}, and we consider only non-contiguous composite
pulses (NCPs) \cite{i18.1}. These pulses, as in Ref. \cite{i10}, must be in
resonance with atomic transitions $\left\vert \mathbf{p}+m\hbar \mathbf{k}%
\right\rangle \leftrightarrow \left\vert \mathbf{p}+\left( m-1\right) \hbar 
\mathbf{k}\right\rangle $, where $m>1$ is an integer, their frequencies
differ from the frequencies of the main pulses, resonant to transitions $%
\left\vert \mathbf{p}\right\rangle \longleftrightarrow \left\vert \mathbf{p}+%
\mathbf{\hbar k}\right\rangle $, and they also differ from each other.
However, the Rabi frequency for all pulses is the same, and the sum of the
pulse durations in a given NCP should be equal to $2\tau $, where $\tau $ is
the duration of the first and third main $\dfrac{\pi }{2}-$ pulses. If the
composite pulse consists of two rectangular pulses, then each of them can
have a duration $\tau $.

\begin{widetext}
\begin{center} 
\begin{figure}[t]
\includegraphics[width=.89\columnwidth]{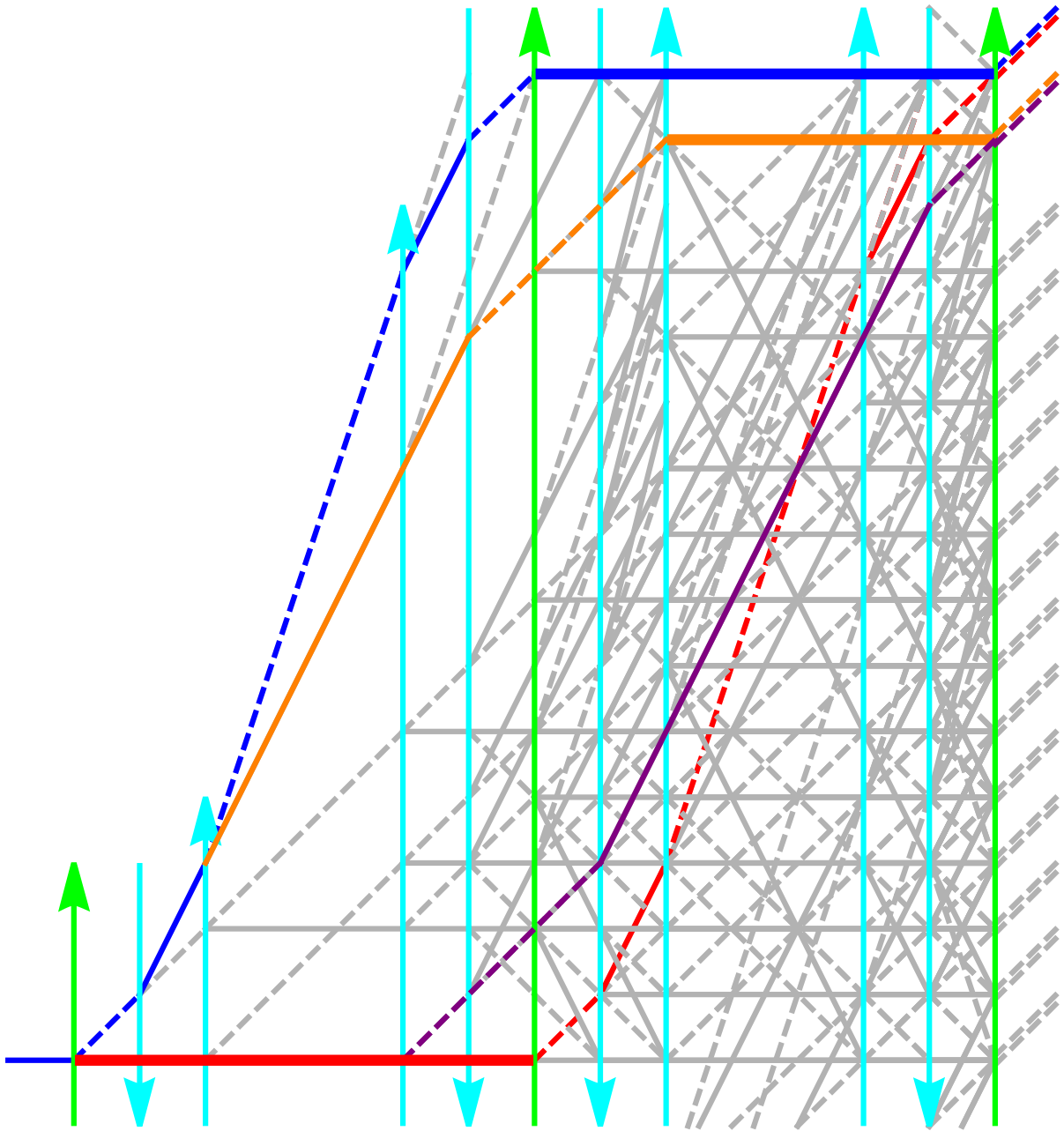}
\caption{Same as in Fig. \protect\ref{f2}, but for $n=2$, when each set of
auxiliary pulses contains two $\protect\pi -$pulses with opposite effective
wave vectors. One included only color output ports in consideration.}
\label{f3}
\end{figure}
\end{center} 
\end{widetext}%

Below we consider the NCP, consisting of three rectangular pulses. If the
durations of the first and third pulses coincide, then obviously the
duration of each of the rectangular pulses can vary in the range $\left[
0,\tau \right] $. Under the action of a composite pulse, the atom performs
Rabi oscillations during each of the pulses, and nutation of atomic
coherence takes place in the time between rectangular pulses. Below we find
such delay between pulses%
\begin{equation}
\tau _{b}\sim \omega _{k}^{-1},  \label{16}
\end{equation}%
under which, owing to a combination of nutation and Rabi oscillations,
requirement \ref{II} is satisfied precisely.

\subsection{Recoil phase}

It is well known that atomic interference is caused by the quantization of
the motion of the atomic center-of-mass. When the incident atomic momentum
state $\left\vert \mathbf{p}\right\rangle $ splits into two states $%
\left\vert \mathbf{p}\right\rangle $ and $\left\vert \mathbf{p}+\mathbf{%
\hbar k}\right\rangle $ after passing through the beam splitter, the
coherence between these states evolves as%
\begin{equation}
\rho \left( \mathbf{p}+\hbar \mathbf{k},\mathbf{p},t\right) \propto \exp
\left( -i\omega _{\mathbf{p}+\hbar \mathbf{k},\mathbf{p}}t\right) ,
\label{17}
\end{equation}%
where the frequency of transition between states 
\begin{subequations}
\label{18}
\begin{eqnarray}
\omega _{\mathbf{p}+\hbar \mathbf{k},\mathbf{p}} &=&\dfrac{1}{2M\hbar }\left[
\left( \mathbf{p}+\hbar \mathbf{k}\right) ^{2}-\mathbf{p}^{2}\right] =\omega
_{D}+\omega _{k},  \label{18a} \\
\omega _{D} &=&\mathbf{k}\cdot \dfrac{\mathbf{p}}{M},  \label{18b}
\end{eqnarray}%
along with the Doppler frequency shift $\omega _{D}$, contains a quantum
term, the recoil frequency (\ref{2}). If $t\sim T$ one can expect that the
AI phase contains Doppler and quantum terms 
\end{subequations}
\begin{subequations}
\label{19}
\begin{eqnarray}
\phi _{D} &\sim &\omega _{D}T,  \label{19a} \\
\phi _{q} &\sim &\omega _{k}T.  \label{19b}
\end{eqnarray}%
Despite this, the phase of the well-known and widely used MZAI does not
contain any quantum term in a uniform gravity field [see Eq. (\ref{5b}) for $%
n=0$]. The reason is that, although quantum corrections affect changes in
the atom's coordinates at the moments of interaction with the second and
third beam splitters, the corresponding quantum terms compensate each other
in a uniform gravity field (see Appendix A in Ref. \cite{i19}). We would
like to emphasize that the derivation of the expression for the AI phase is
purely quantum (see examples of this derivation in Refs. \cite{i20,i6,i19}),
but the result of these derivations, Eq. (\ref{5b}), is purely classical.
The absence of a quantum phase (\ref{19b}) allows one to be in doubt that
the MZAI is caused by the matter waves interference (a sentiment held by the
present author). Examples of the derivation of the expression for the phase
without using the atomic center-of-mass motion quantization can be found in
preprint \cite{i20.1}.

The quantum contribution arises in the rotating reference frame \cite{i21},
or in the non-uniform gravity field in the presence of the gravity-gradient
tensor \cite{i22,i23}, or in the presence of the gravity-gradient tensor of
the second order \cite{i24}, or in a strongly inhomogeneous source mass
field \cite{i24}. The quantum term in the gravity-gradient field of the
source mass was observed in \cite{i25} using the LMT method. In all these
cases, the magnitude of the quantum terms is small compared to the phase $%
\phi _{q}$ in Eq. (\ref{19b}). We predict here that the situation may change
dramatically in the presence of auxiliary pulses. If the ultimate goal is
not to create a high-precision gravimeter, but, as in \cite{i25}, to observe
the quantum term, and the timing of these pulses is comparable to the
interrogation time $T$, then the quantum term turns out to be large and
grows as $n/\ln n$ with increasing momentum transfer.

It should be noted, however, that the quantum term does not arise if, as in
the experiments in \cite{i10,i14,i16}, the auxiliary pulses are equidistant
in time. With nonequidistant timing, the quantum term does not depend on the
shape and duration of the pulse, it is the same for the rectangular pulses
considered here, as well as in the Bragg regime \cite{i10}. However, for
SLMT in the Raman-Nath approximation (\ref{3.1}), for the both Raman beam
splitters and atomic clocks \cite{i14}, the quantum term must be
recalculated.
\begin{figure}[t]
\includegraphics[width=8cm]{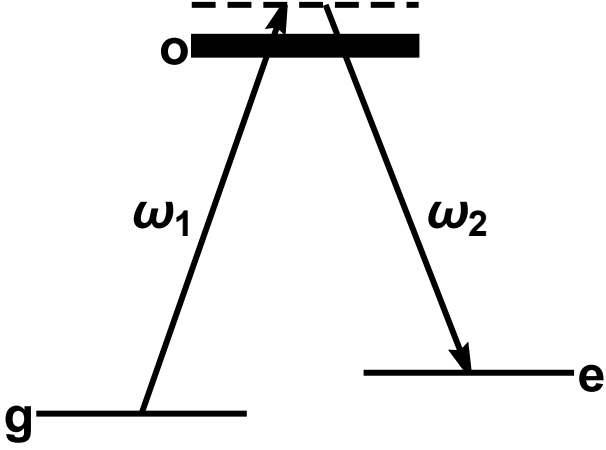}
\caption{Atomic level diagram}
\label{f3.5}
\end{figure}

\subsection{The article structure}

In this article, we use the Schr\"{o}dinger equation in momentum space, in
contrast to the same approach in preprints \cite{i19,i26}, here we take into
account the motion of the atom during the time of interaction with the pulse.

The overall plan of the paper is as follows. In the next section, we use the
solutions of the Schr\"{o}dinger equation for the interaction of an atom
with a rectangular pulse to determine the parameters of the NCPs, consisting
of one, two and three rectangular pulses. In Sec. \ref{sIII} we calculate
the MZAI phase. In Sec. \ref{sIV} we consider 2 types of quantum parts of
MZAI phases. In Sec. \ref{sV}, an expression for the gravity part of the
MZAI phase is obtained. We calculate the part of the gravity phase,
independent of the duration of the main and auxiliary pulses, and
corrections linear in these durations.

\onecolumngrid%

\section{Main relations.}

Let us consider the interaction of a three-level atom with the pulse of the
field of two traveling waves resonant to adjacent atomic transitions 
\end{subequations}
\begin{equation}
\mathbf{E}\left( \mathbf{x},t\right) =\left\{ \mathbf{E}_{1}\exp \left[
i\left( \mathbf{q}_{1}\cdot \mathbf{x}-\omega _{1}t-\phi _{1}\left( t\right)
\right) \right] +\mathbf{E}_{2}\exp \left[ i\left( \mathbf{q}_{2}\cdot 
\mathbf{x}-\omega _{2}t-\phi _{2}\left( t\right) \right) \right] \right\}
f\left( t\right) +c.c,  \label{20}
\end{equation}%
where $E_{i},$ $\mathbf{q}_{i},\omega _{i}$ and $\phi _{i}\left( t\right) $
are the amplitudes, wave vectors, frequencies and phases of the waves,
respectively, and $f\left( t\right) $ is the shape of the pulse acting at
the moment $T$ and having a duration $\tau $. We assume that the two atomic
states $\left\vert g\right\rangle $ and $\left\vert e\right\rangle $ are
sublevels of the hyperfine structure of the ground-state manifold of the
atom, while the third state $\left\vert o\right\rangle $ is a sublevel of
the excited-state manifold; the fields $\mathbf{E}_{1}$and $\mathbf{E}_{2}$
are resonant to the atomic transitions $g\rightarrow o$ and $e\rightarrow o$
(see Fig. \ref{f3.5}). The location of the sublevels $\left\vert
g\right\rangle $ and $\left\vert e\right\rangle $ on the atomic energy
diagram is not important. However, we consider sublevels $\left\vert
g\right\rangle $ and $\left\vert e\right\rangle $ to be ground and excited.

The Hamiltonian of the interaction of an atom with a field is%
\begin{equation}
H=\dfrac{p^{2}}{2M}-M\mathbf{g\cdot x}+\dfrac{\hbar }{2}\left\{ \Omega
_{1}\exp \left[ i\left( \mathbf{q}_{1}\cdot \mathbf{x}-\Delta _{1}t-\phi
_{1}\left( t\right) \right) \right] \left\vert o\right\rangle \left\langle
g\right\vert +\Omega _{2}\exp \left[ i\left( \mathbf{q}_{2}\cdot \mathbf{x}%
-\Delta _{2}t-\phi _{2}\left( t\right) \right) \right] \left\vert
o\right\rangle \left\langle e\right\vert +H.c.\right\} f\left( t\right) ,
\label{22}
\end{equation}%
where $\mathbf{p}$ is the momentum of the atom; and $\mathbf{g}$ is the
gravity field; 
\begin{subequations}
\label{23}
\begin{eqnarray}
\Omega _{1} &\equiv &-2\dfrac{\mathbf{d}_{og}\mathbf{\cdot E}_{1}}{\hbar },
\label{23a} \\
\Omega _{2} &\equiv &-2\dfrac{\mathbf{d}_{oe}\mathbf{\cdot E}_{2}}{\hbar }
\label{23b}
\end{eqnarray}%
are the Rabi frequencies of atomic transitions, with $\mathbf{d}$ the dipole
moment operator; and 
\end{subequations}
\begin{subequations}
\label{24}
\begin{eqnarray}
\Delta _{1} &=&\omega _{1}-\omega _{og},  \label{24a} \\
\Delta _{2} &=&\omega _{2}-\omega _{oe}  \label{24b}
\end{eqnarray}%
are frequency detunings of the fields. The amplitudes of atomic levels
evolve as 
\end{subequations}
\begin{subequations}
\label{25}
\begin{eqnarray}
i\partial _{t}\tilde{a}\left( e,\mathbf{p},t\right)  &=&\left( \dfrac{p^{2}}{%
2M\hbar }-iM\mathbf{g\cdot }\partial _{\mathbf{p}}\right) \tilde{a}\left( e,%
\mathbf{p},t\right) +\dfrac{\Omega _{2}^{\ast }}{2}\exp \left[ i\left(
\Delta _{2}t+\phi _{2}\left( t\right) \right) \right] f\left( t\right) 
\tilde{a}\left( o,\mathbf{p}+\hbar \mathbf{q}_{2},t\right) ,  \label{25a} \\
i\partial _{t}\tilde{a}\left( g,\mathbf{p},t\right)  &=&\left( \dfrac{p^{2}}{%
2M\hbar }-iM\mathbf{g\cdot }\partial _{\mathbf{p}}\right) \tilde{a}\left( g,%
\mathbf{p},t\right) +\dfrac{\Omega _{1}^{\ast }}{2}\exp \left[ i\left(
\Delta _{1}t+\phi _{1}\left( t\right) \right) \right] f\left( t\right) 
\tilde{a}\left( o,\mathbf{p}+\hbar \mathbf{q}_{1},t\right) ,  \label{25b} \\
i\partial _{t}\tilde{a}\left( o,\mathbf{p},t\right)  &=&\left( \dfrac{p^{2}}{%
2M\hbar }-iM\mathbf{g\cdot }\partial _{\mathbf{p}}\right) \tilde{a}\left( o,%
\mathbf{p},t\right) +\dfrac{\Omega _{1}}{2}\exp \left[ -i\left( \Delta
_{1}t+\phi _{1}\left( t\right) \right) \right] f\left( t\right) \tilde{a}%
\left( g,\mathbf{p}-\hbar \mathbf{q}_{1},t\right)   \notag \\
&&+\dfrac{\Omega _{2}}{2}\exp \left[ -i\left( \Delta _{2}t+\phi _{2}\left(
t\right) \right) \right] f\left( t\right) \tilde{a}\left( e,\mathbf{p}-\hbar 
\mathbf{q}_{2},t\right) .  \label{25c}
\end{eqnarray}%
One assumes that the frequency detuning is large enough 
\end{subequations}
\begin{subequations}
\label{26}
\begin{eqnarray}
\Delta _{1} &\approx &\Delta _{2}\approx \Delta ,  \label{26a} \\
\left\vert \Delta \right\vert  &\gg &\max \left\{ \left\vert \tilde{\delta}%
\right\vert ,\tau _{f}^{-1},\left\vert \dot{\phi}_{i}\right\vert ,\left\vert 
\mathbf{q}_{i}\cdot \mathbf{g}\right\vert T,\left\vert \omega
_{D}\right\vert ,\omega _{k}\right\} ,  \label{26b}
\end{eqnarray}%
where 
\end{subequations}
\begin{equation}
\tilde{\delta}=\Delta _{1}-\Delta _{2}  \label{27}
\end{equation}%
is a Raman detuning, and $\tau _{f}$ is the duration of the forward and
backward fronts of the pulse. At this assumption one finds the expression
for the level $\left\vert o\right\rangle $ amplitude \cite{i26.1}%
\begin{equation}
\tilde{a}\left( o,\mathbf{p},t\right) =\dfrac{f\left( t\right) }{2\Delta }%
\left\{ \Omega _{1}\exp \left[ -i\left( \Delta _{1}t+\phi _{1}\left(
t\right) \right) \right] \tilde{a}\left( g,\mathbf{p}-\hbar \mathbf{q}%
_{1},t\right) +\Omega _{2}\exp \left[ -i\left( \Delta _{2}t+\phi _{2}\left(
t\right) \right) \right] \tilde{a}\left( e,\mathbf{p}-\hbar \mathbf{q}%
_{2},t\right) \right\} .  \label{28}
\end{equation}%
Then for the amplitudes of the ground state sublevels one gets 
\begin{subequations}
\label{29}
\begin{eqnarray}
i\partial _{t}\tilde{a}\left( e,\mathbf{p},t\right)  &=&\left( \dfrac{p^{2}}{%
2M\hbar }-iM\mathbf{g\cdot }\partial _{\mathbf{p}}+\dfrac{\left\vert \Omega
_{2}\right\vert ^{2}}{4\Delta }\right) \tilde{a}\left( e,\mathbf{p},t\right)
+\dfrac{\Omega }{2}\exp \left[ -i\left( \tilde{\delta}t+\phi \left( t\right)
\right) \right] f^{2}\left( t\right) \tilde{a}\left( g,\mathbf{p}-\hbar 
\mathbf{k},t\right) ,  \label{29a} \\
i\partial _{t}\tilde{a}\left( g,\mathbf{p},t\right)  &=&\left( \dfrac{p^{2}}{%
2M\hbar }-iM\mathbf{g\cdot }\partial _{\mathbf{p}}+\dfrac{\left\vert \Omega
_{1}\right\vert ^{2}}{4\Delta }\right) \tilde{a}\left( g,\mathbf{p},t\right)
+\dfrac{\Omega ^{\ast }}{2}\exp \left[ i\left( \tilde{\delta}t+\phi \left(
t\right) \right) \right] f^{2}\left( t\right) \tilde{a}\left( e,\mathbf{p}%
+\hbar \mathbf{k},t\right) ,  \label{29b}
\end{eqnarray}%
where 
\end{subequations}
\begin{subequations}
\label{30}
\begin{eqnarray}
\mathbf{k} &=&\mathbf{q}_{1}-\mathbf{q}_{2},  \label{30a} \\
\Omega  &=&\dfrac{\Omega _{1}\Omega _{2}^{\ast }}{2\Delta },  \label{30b} \\
\phi \left( t\right)  &=&\phi _{1}\left( t\right) -\phi _{2}\left( t\right) 
\label{30c}
\end{eqnarray}%
are the effective wave vector, the Rabi frequency, and the phase of the
Raman beam splitter, respectively. Eliminating the ac-Stark shift by a
transformation to a rotating interaction picture, i.e. by introducing the
amplitudes 
\end{subequations}
\begin{equation}
\left( 
\begin{array}{c}
\tilde{a}\left( e,\mathbf{p},t\right)  \\ 
\tilde{a}\left( g,\mathbf{p},t\right) 
\end{array}%
\right) =\exp \left[ -\dfrac{i}{4\Delta }\int^{t}dt^{\prime }\left( 
\begin{array}{c}
\left\vert \Omega _{2}\right\vert ^{2} \\ 
\left\vert \Omega _{1}\right\vert ^{2}%
\end{array}%
\right) f^{2}\left( t^{\prime }\right) \right] \left( 
\begin{array}{c}
a\left( e,\mathbf{p},t\right)  \\ 
a\left( g,\mathbf{p},t\right) 
\end{array}%
\right)   \label{31}
\end{equation}%
one obtains 
\begin{subequations}
\label{32}
\begin{eqnarray}
i\left( \partial _{t}+M\mathbf{g\cdot }\partial _{\mathbf{p}}+i\dfrac{p^{2}}{%
2M\hbar }\right) a\left( e,\mathbf{p},t\right)  &=&\dfrac{\Omega }{2}%
f^{2}\left( t\right) \exp \left[ -i\delta t-i\phi \left( t\right) \right]
a\left( g,\mathbf{p}-\hbar \mathbf{k},t\right) ,  \label{32a} \\
i\left( \partial _{t}+M\mathbf{g\cdot }\partial _{\mathbf{p}}+i\dfrac{p^{2}}{%
2M\hbar }\right) a\left( g,\mathbf{p},t\right)  &=&\dfrac{\Omega ^{\ast }}{2}%
f^{2}\left( t\right) \exp \left( i\delta t+i\phi \left( t\right) \right)
a\left( e,\mathbf{p}+\hbar \mathbf{k},t\right) ,  \label{32b}
\end{eqnarray}%
where 
\end{subequations}
\begin{subequations}
\label{33}
\begin{eqnarray}
\delta  &=&\tilde{\delta}-\delta _{S},  \label{33a} \\
\delta _{S} &=&\dfrac{\left\vert \Omega _{2}\right\vert ^{2}-\left\vert
\Omega _{1}\right\vert ^{2}}{4\Delta }  \label{33b}
\end{eqnarray}%
is the ac-Stark shift of the Raman line. The presence of an ac-Stark shift
would significantly complicate the present study. We will assume that this
shift is absent. Experimental methods for eliminating the ac-Stark shift
were developed in Refs. \cite{i26.3,i26.4}. If, for example, the Raman pulse
has an area $\dfrac{\pi }{2}$, then from the equation $\delta _{S}=0$ and
Eq. (\ref{30b}) one has 
\end{subequations}
\begin{equation}
\left\vert \Omega _{1}\right\vert =\sqrt{\dfrac{\pi \left\vert \Delta
\right\vert }{\tau }}.  \label{34}
\end{equation}%
At typical values of $\Delta =2\pi \times 10$GHz$\ $and $\tau =30~\mu $s,
the parameter%
\begin{equation}
\left\vert \Omega _{1}\right\vert /\Delta \approx 7\times 10^{-4}
\label{34.1}
\end{equation}%
is small enough to neglect the ac-Stark splitting of optical transitions $%
e\rightarrow o$ and $g\rightarrow o$\cite{i27}.

In an accelerated frame 
\begin{equation}
\mathbf{p}=\mathbf{P}+M\mathbf{g}t,  \label{35}
\end{equation}%
using the atom's initial momentum $\mathbf{P}$ as the independent variable,
one gets 
\begin{subequations}
\label{36}
\begin{eqnarray}
i\left( \dfrac{d}{dt}+i\dfrac{\left( \mathbf{P}+M\mathbf{g}t\right) ^{2}}{%
2M\hbar }\right) a\left( e,\mathbf{P},t\right) &=&\dfrac{\Omega }{2}%
f^{2}\left( t\right) \exp \left[ -i\delta t-i\phi \left( t\right) \right]
a\left( g,\mathbf{P}-\hbar \mathbf{k},t\right) ,  \label{36a} \\
i\left( \dfrac{d}{dt}+i\dfrac{\left( \mathbf{P}+M\mathbf{g}t\right) ^{2}}{%
2M\hbar }\right) a\left( g,\mathbf{P},t\right) &=&\dfrac{\Omega ^{\ast }}{2}%
f^{2}\left( t\right) \exp \left[ i\delta t+i\phi \left( t\right) \right]
a\left( e,\mathbf{P}+\hbar \mathbf{k},t\right) .  \label{36b}
\end{eqnarray}%
Then one finds that in the interaction representation 
\end{subequations}
\begin{equation}
a\left( n,\mathbf{P},t\right) =\exp \left[ -i\int_{0}^{t}dt^{\prime }\dfrac{%
\left( \mathbf{P}+M\mathbf{g}t^{\prime }\right) ^{2}}{2M\hbar }\right]
c\left( n,\mathbf{P},t\right)  \label{36.1}
\end{equation}%
state vector%
\begin{equation}
\underline{c}\left( \mathbf{P},t\right) =\left( 
\begin{array}{c}
c\left( e,\mathbf{P}+\dfrac{\hbar \mathbf{k}}{2},t\right) \\ 
c\left( g,\mathbf{P}-\dfrac{\hbar \mathbf{k}}{2},t\right)%
\end{array}%
\right)  \label{37}
\end{equation}%
evolves as%
\begin{equation}
i\underline{\dot{c}}=\dfrac{f^{2}\left( t\right) }{2}\left( 
\begin{array}{cc}
0 & \Omega \exp \left\{ -i\left[ \delta t+\phi \left( t\right) -\mathbf{%
k\cdot }\dfrac{\mathbf{P}}{M}t-\dfrac{1}{2}\mathbf{k\cdot g}t^{2}\right]
\right\} \\ 
\Omega ^{\ast }\exp \left\{ i\left[ \delta t+\phi \left( t\right) -\mathbf{%
k\cdot }\dfrac{\mathbf{P}}{M}t-\dfrac{1}{2}\mathbf{k\cdot g}t^{2}\right]
\right\} & 0%
\end{array}%
\right) \underline{c}.  \label{38}
\end{equation}%
One sees that in the accelerated frame the amplitude of the atomic state
remains unchanged outside the field pulse%
\begin{equation}
\underline{c}\left( \mathbf{P},t\right) =\text{const at }t<T\text{ or }%
t>T+\tau  \label{39}
\end{equation}%
If one chirps the field frequency linearly, i.e., if%
\begin{equation}
\phi \left( t\right) =\phi +\alpha t^{2}/2,  \label{40}
\end{equation}%
and if the chirping rate $\alpha $ is close to $\mathbf{k}\cdot \mathbf{g},$%
\begin{equation}
\left\vert \alpha -\mathbf{k\cdot g}\right\vert \tau ^{2}\ll 1,  \label{41}
\end{equation}%
then, considering Eq. (\ref{38}) when%
\begin{equation}
t=T+\varepsilon ,  \label{42}
\end{equation}%
where%
\begin{equation}
\varepsilon \sim \tau \ll T,  \label{43}
\end{equation}%
and neglecting terms quadratic in $\varepsilon $ in the phase factors in Eq.
(\ref{38}), one arrives at the well-known equation for the amplitudes of a
two-level atom interacting with the pulse of the field of arbitrary shape
with constant frequency and phase%
\begin{equation}
i\dfrac{d\underline{c}}{d\varepsilon }=\dfrac{f^{2}\left( T+\varepsilon
\right) }{2}\left( 
\begin{array}{cc}
0 & \Omega \exp \left\{ -i\left[ \nu \varepsilon +\phi \left( \mathbf{P}%
\right) \right] \right\} \\ 
\Omega ^{\ast }\exp \left\{ i\left[ \nu \varepsilon +\phi \left( \mathbf{P}%
\right) \right] \right\} & 0%
\end{array}%
\right) \underline{c},  \label{44}
\end{equation}%
where 
\begin{subequations}
\label{45}
\begin{eqnarray}
\nu &\equiv &\delta \left( \mathbf{P}\right) =\nu ^{\left( 0\right) }+\nu
^{\left( 1\right) },  \label{45a} \\
\nu ^{\left( 0\right) } &\equiv &\delta ^{\left( 0\right) }\left( \mathbf{P}%
\right) =\delta -\mathbf{k\cdot }\dfrac{\mathbf{P}}{M},  \label{45b} \\
\nu ^{\left( 1\right) } &=&-\left( \mathbf{k\cdot g}-\alpha \right) T,
\label{45c} \\
\phi \left( \mathbf{P}\right) &=&\phi +\nu ^{\left( 0\right) }T-\dfrac{1}{2}%
\left( \mathbf{k\cdot g}-\alpha \right) T^{2}.  \label{45d}
\end{eqnarray}%
After unitary transformation 
\end{subequations}
\begin{equation}
\underline{c}=\underline{U}_{\phi }\underline{\tilde{b}},  \label{46}
\end{equation}%
where the matrix $U_{\phi }$ is given by%
\begin{equation}
\underline{U}_{\phi }=\left( 
\begin{array}{cc}
\exp \left[ -\dfrac{i}{2}\phi \left( \mathbf{P}\right) \right] & 0 \\ 
0 & \exp \left[ \dfrac{i}{2}\phi \left( \mathbf{P}\right) \right]%
\end{array}%
\right) ,  \label{47}
\end{equation}%
one arrives at the phase-independent equation%
\begin{equation}
i\dfrac{d\underline{\tilde{b}}}{d\varepsilon }=\dfrac{f^{2}\left(
T+\varepsilon \right) }{2}\left( 
\begin{array}{cc}
0 & \Omega \exp \left( -i\nu \varepsilon \right) \\ 
\Omega ^{\ast }\exp \left( i\nu \varepsilon \right) & 0%
\end{array}%
\right) \underline{\tilde{b}}.  \label{48}
\end{equation}%
In the AI phase, the factor $\exp \left[ \pm \dfrac{i}{2}\phi \left( \mathbf{%
P}\right) \right] $ will be responsible for the Doppler and quantum phases (%
\ref{19}), for the gravity phase (\ref{5b}), as well as for the Ramsey phase%
\begin{equation}
\phi _{R}\sim \delta T.  \label{49}
\end{equation}%
One might conclude that the results to be obtained below for the quantum,
gravity and Ramsey phases are independent of the pulse shape $f\left(
t\right) $ and do not change from the Raman-Nath regime (\ref{3.1}) to the
Bragg regime (\ref{1}).

Consider now a NCP consisting of $\ell $ time-separated rectangular pulses%
\begin{equation}
f\left( T+\varepsilon \right) =\left\{ 
\begin{array}{l}
1\text{\quad for }\varepsilon \in \dbigcup\limits_{m=1}^{\ell }\left[ \tau
_{0,m},\tau _{0,m}+\tau _{m}\right] \\ 
0\text{\quad in other cases,}%
\end{array}%
\right.  \label{50}
\end{equation}%
where one turns on a pulse of duration $\tau _{m}$ at time $T+\tau _{0,m}$,
so that%
\begin{equation}
\tau _{0,m}=\dsum_{m^{\prime }=1}^{m-1}\left( \tau _{m^{\prime }}+\tau
_{bm^{\prime }}\right) ,  \label{51}
\end{equation}%
where $\tau _{bm}$ is the delay time between adjacent pulses $m$ and $m+1$.
After the next unitary transformation 
\begin{subequations}
\label{52}
\begin{eqnarray}
\underline{\tilde{b}} &=&\underline{U}_{\delta }\underline{b},  \label{52a}
\\
U_{\delta } &=&\left( 
\begin{array}{cc}
\exp \left( -\dfrac{i}{2}\nu \varepsilon \right) & 0 \\ 
0 & \exp \left( \dfrac{i}{2}\nu \varepsilon \right)%
\end{array}%
\right)  \label{52b}
\end{eqnarray}%
one finds 
\end{subequations}
\begin{subequations}
\label{53}
\begin{eqnarray}
i\dfrac{d\underline{b}}{d\varepsilon } &=&\underline{h}\underline{b},
\label{53a} \\
\underline{h} &=&\dfrac{1}{2}\left( 
\begin{array}{cc}
-\nu & \Omega \\ 
\Omega ^{\ast } & +\nu%
\end{array}%
\right)  \label{53b}
\end{eqnarray}%
The solution of this equation is well known. One can achieve it using
composite rotation matrices. Alternatively, one can represent the $h-$matrix
as 
\end{subequations}
\begin{equation*}
\underline{h}=\mathbf{h\cdot }\underline{\mathbf{\sigma }},
\end{equation*}%
where 
\begin{equation*}
\mathbf{h=}\dfrac{1}{2}\left\{ \func{Re}\Omega ,-\func{Im}\Omega ,-\nu
\right\}
\end{equation*}%
and $\underline{\mathbf{\sigma }}=\left\{ \underline{\sigma }_{1},\underline{%
\sigma }_{2},\underline{\sigma }_{3}\right\} ,$~with $\underline{\sigma }%
_{i} $ the Pauli matrix, and for the $s-$matrix $\underline{\sigma }%
=epx\left( -i\underline{h}t\right) $ use the expression \cite{i26.2} 
\begin{equation}
\underline{f}\left( \mathbf{h\cdot \underline{\sigma }}\right) =\dfrac{1}{2}%
\left\{ f\left( h\right) +f\left( -h\right) +\dfrac{\mathbf{h\cdot 
\underline{\sigma }}}{h}\left[ f\left( h\right) -f\left( -h\right) \right]
\right\} .  \label{53.1}
\end{equation}%
Immediately after the action of the pulse $m$, the wave function of the atom
is 
\begin{subequations}
\label{54}
\begin{eqnarray}
\underline{b}\left( T+\tau _{0,m}+\tau _{m}\right) &=&\underline{\sigma }%
\left( \tau _{m}\right) \underline{b}\left( T+\tau _{0,m}\right) ,
\label{54a} \\
\underline{\sigma }\left( \tau \right) &=&\left( 
\begin{array}{cc}
s_{d}\left( \tau \right) & -is_{a}\left( \tau \right) \\ 
-is_{a}^{\ast }\left( \tau \right) & s_{d}^{\ast }\left( \tau \right)%
\end{array}%
\right) ,  \label{54b} \\
s_{d}\left( \tau \right) &=&\cos \dfrac{\Omega _{r}\tau }{2}+i\dfrac{\nu }{%
\Omega _{r}}\sin \dfrac{\Omega _{r}\tau }{2},  \label{54c} \\
s_{a}\left( \tau \right) &=&\dfrac{\Omega }{\Omega _{r}}\sin \dfrac{\Omega
_{r}\tau }{2},  \label{54d} \\
\Omega _{r} &=&\sqrt{\left\vert \Omega \right\vert ^{2}+\nu ^{2}}.
\label{54e}
\end{eqnarray}%
Hence, for $\Omega =0$, corresponding to the time between pulses $\tau _{bm}$%
, the $s-$matrix is 
\end{subequations}
\begin{equation}
\underline{s}_{bm}=\left( 
\begin{array}{cc}
\exp \left( \dfrac{i}{2}\nu \tau _{bm}\right) & 0 \\ 
0 & \exp \left( -\dfrac{i}{2}\nu \tau _{bm}\right)%
\end{array}%
\right) .  \label{55}
\end{equation}%
The numbering of auxiliary NCPs $\left\{ \zeta ,\beta \right\} $ will be
introduced below in Sec. \ref{sIII}, where one verifies that either the
pulse duration at $\ell =1$ or the distance between pulses at $\ell >1$
depends on the Raman detuning $\nu $ on the nonresonant branch of the recoil
diagram so that the total time of action of this NCP is a function of four
parameters $\left\{ \nu ,\zeta ,\beta ,\ell \right\} $, for which one has%
\begin{equation}
\tau \left( \nu ,\zeta ,\beta ,\ell \right) =\tau _{\ell }+\dsum_{m=1}^{\ell
-1}\left( \tau _{m}+\tau _{bm}\right) ,  \label{56}
\end{equation}%
For the $s-$matrix one has%
\begin{equation}
\underline{s}=\underline{\sigma }\left( \tau _{\ell }\right) \underline{s}%
_{b\left( \ell -1\right) }\underline{\sigma }\left( \tau _{\ell -1}\right)
\cdots \underline{s}_{b1}\underline{\sigma }\left( \tau _{1}\right) .
\label{57}
\end{equation}%
Now, returning to the state vector (\ref{37}) and to the lab frame, 
\begin{subequations}
\label{58}
\begin{eqnarray}
\mathbf{P} &=&\mathbf{p}_{+,T+\tau \left( \nu ,\zeta ,\beta ,\ell \right) },
\label{58a} \\
\mathbf{p}_{t} &\equiv &\mathbf{p}-M\mathbf{g}t,  \label{58b}
\end{eqnarray}%
one arrives at the next result 
\end{subequations}
\begin{subequations}
\label{59}
\begin{gather}
c\left( e,\mathbf{p}_{+},T+\tau \left( \nu ,\zeta ,\beta ,\ell \right)
\right) =S_{ee}\left( \mathbf{p}_{+,T+\tau \left( \nu ,\zeta ,\beta ,\ell
\right) }-\dfrac{\hbar \mathbf{k}}{2}\right) c\left( e,\mathbf{p}%
_{-},T\right) +S_{eg}\left( \mathbf{p}_{+,T+\tau \left( \nu ,\zeta ,\beta
,\ell \right) }-\dfrac{\hbar \mathbf{k}}{2}\right) c\left( g,\mathbf{p}%
_{-}-\hbar \mathbf{k},T\right) ,  \label{59a} \\
c\left( g,\mathbf{p}_{+},T+\tau \left( \nu ,\zeta ,\beta ,\ell \right)
\right) =S_{ge}\left( \mathbf{p}_{+,T+\tau \left( \nu ,\zeta ,\beta ,\ell
\right) }+\dfrac{\hbar \mathbf{k}}{2}\right) c\left( e,\mathbf{p}_{-}+\hbar 
\mathbf{k},T\right) +S_{gg}\left( \mathbf{p}_{+,T+\tau \left( \nu ,\zeta
,\beta ,\ell \right) }+\dfrac{\hbar \mathbf{k}}{2}\right) c\left( g,\mathbf{p%
}_{-},T\right) ,  \label{59b}
\end{gather}%
where 
\end{subequations}
\begin{subequations}
\label{60}
\begin{eqnarray}
S\left( \mathbf{P}\right) &=&\left( 
\begin{array}{cc}
\hat{s}_{ee} & \exp \left[ -i\phi \left( \mathbf{P}\right) \right] \hat{s}%
_{eg} \\ 
\exp \left[ i\phi \left( \mathbf{P}\right) \right] \hat{s}_{ge} & \hat{s}%
_{gg}%
\end{array}%
\right) ,  \label{60a} \\
\underline{\hat{s}} &=&\left( 
\begin{array}{cc}
\exp \left[ -\dfrac{i}{2}\nu \tau \left( \nu ,\zeta ,\beta ,\ell \right) %
\right] s_{ee} & \exp \left[ -\dfrac{i}{2}\nu \tau \left( \nu ,\zeta ,\beta
,\ell \right) \right] s_{eg} \\ 
\exp \left[ \dfrac{i}{2}\nu \tau \left( \nu ,\zeta ,\beta ,\ell \right) %
\right] s_{ge} & \exp \left[ \dfrac{i}{2}\nu \tau \left( \nu ,\zeta ,\beta
,\ell \right) \right] s_{gg}%
\end{array}%
\right) .  \label{60b}
\end{eqnarray}%
where $s_{\alpha \beta }$ is a matrix element of the $s-$matrix (\ref{57}).

%
The momentum of the atom changes due to the transfer of the momentum of the
photon $\pm \hbar \mathbf{k}$ and under the action of gravitation $\mathbf{g}
$. The change in the momentum of the atom under the action of a uniform
gravitational field does not depend on the initial value of the momentum
[see Eq. (\ref{58b})] and therefore does not depend on the momentum of the
photon transferred to the atom. This allows one to represent the momenta of
an atom as 
\end{subequations}
\begin{equation}
\mathbf{p}_{\pm }=\mathbf{p}_{\pm }^{\prime }+\mathfrak{N\hbar }\mathbf{k},
\label{60.1}
\end{equation}%
where $\mathfrak{N}$ is the total number of photon momenta transferred to
the atom at a given point in the recoil diagram. Below we omit the prime in
the expressions for momenta. If the effective wave vector of a given NCP is
equal to $\pm \mathbf{k}$, then, keeping in mind the redefinition (\ref{60.1}%
), one obtains, instead of Eqs. (\ref{59}), 
\begin{subequations}
\label{60.2}
\begin{eqnarray}
c\left( e,\mathbf{p}_{+}+\mathfrak{N}\hbar \mathbf{k},T+\tau \left( \nu
,\zeta ,\beta ,\ell \right) \right) &=&S_{ee}\left( \mathbf{p}_{+,T+\tau
\left( \nu ,\zeta ,\beta ,\ell \right) }+\left( 2\mathfrak{N}\mp 1\right) 
\dfrac{\hbar \mathbf{k}}{2}\right) c\left( e,\mathbf{p}_{-}+\mathfrak{N}%
\hbar \mathbf{k},T\right)  \notag \\
&&+S_{eg}\left( \mathbf{p}_{+,T+\tau \left( \nu ,\zeta ,\beta ,\ell \right)
}+\left( 2\mathfrak{N}\mp 1\right) \dfrac{\hbar \mathbf{k}}{2}\right)
c\left( g,\mathbf{p}_{-}+\left( \mathfrak{N}\mp 1\right) \hbar \mathbf{k}%
,T\right) ,  \label{60.2a} \\
c\left( g,\mathbf{p}_{+}+\mathfrak{N}\hbar \mathbf{k},T+\tau \left( \nu
,\zeta ,\beta ,\ell \right) \right) &=&S_{ge}\left( \mathbf{p}_{+,T+\tau
\left( \nu ,\zeta ,\beta ,\ell \right) }+\left( 2\mathfrak{N}\pm 1\right) 
\dfrac{\hbar \mathbf{k}}{2}\right) c\left( e,\mathbf{p}_{-}+\left( \mathfrak{%
N}\pm 1\right) \hbar \mathbf{k},T\right)  \notag \\
&&+S_{gg}\left( \mathbf{p}_{+,T+\tau \left( \nu ,\zeta ,\beta ,\ell \right)
}+\left( 2\mathfrak{N}\pm 1\right) \dfrac{\hbar \mathbf{k}}{2}\right)
c\left( g,\mathbf{p}_{-}+\mathfrak{N}\hbar \mathbf{k},T\right) .
\label{60.2b}
\end{eqnarray}%
%

The independent variable is the momentum of the atom after interaction with
the NCP, $\mathbf{p}_{+}$. Then from Eqs. (\ref{58}) it follows that the
momentum of the atom before this interaction is 
\end{subequations}
\begin{equation}
\mathbf{p}_{-}=\mathbf{p}_{+,\tau \left( \nu ,\zeta ,\beta ,\ell \right) }.
\label{61}
\end{equation}%
%
One thus takes into account that during the interaction with the NCP, the
atom was accelerated, i.e. before interaction with a given NCP having a
duration $\tau $, the momentum of the atom $\mathbf{p}_{-}$ was smaller by $M%
\mathbf{g\tau }$ than the momentum after the interaction $\mathbf{p}_{+}$.%
%

If the NCP $\left\{ \zeta ,\beta \right\} $ follows the NCP $\left\{ \zeta
^{\prime },\beta ^{\prime }\right\} $, then it follows from Eq. (\ref{39})
that 
\begin{subequations}
\label{62}
\begin{eqnarray}
\underline{c}\left( \mathbf{p}_{-}^{\left( \zeta ,\beta \right) },T_{\zeta
,\beta }\right) &=&\underline{c}\left( \mathbf{p}_{+}^{\left( \zeta ^{\prime
},\beta ^{\prime }\right) },T_{\zeta ^{\prime },\beta ^{\prime }}+\tau
\left( \nu ^{\prime },\zeta ^{\prime },\beta ^{\prime },\ell ^{\prime
}\right) \right) ,  \label{62a} \\
\mathbf{p}_{+}^{\left( \zeta ^{\prime },\beta ^{\prime }\right) } &=&\mathbf{%
p}_{-,T_{\zeta ,\beta }-T_{\zeta ^{\prime },\beta ^{\prime }}-\tau \left(
\nu ^{\prime },\zeta ^{\prime },\beta ^{\prime },\ell ^{\prime }\right)
}^{\left( \zeta ,\beta \right) },  \label{62b}
\end{eqnarray}%
where $T_{\zeta ,\beta }$ is the moment of action of the NCP $\left\{ \zeta
,\beta \right\} $, $\mathbf{p}_{\pm }^{\left( \zeta ,\beta \right) }$ is the
atomic momentum before and after the action of this NCP, and $\tau \left(
\nu ^{\prime },\zeta ^{\prime },\beta ^{\prime },\ell ^{\prime }\right) $ is
the duration (\ref{56}) of the NCP $\left\{ \zeta ^{\prime },\beta ^{\prime
}\right\} $. 
%
Equations (\ref{62}) mean that atomic state \underline{$c$} stays unchanged
between NCPs and only atomic momentum changes owing to gravity. 
%
Knowing $\mathbf{p}_{+}$, one can restore the momenta of atoms before and
after the action of all preceding NCPs applying consequently Eqs. (\ref{61}, %
\ref{62b}) .

We have not been able to construct an NCP that satisfies requirement \ref{II}
for an arbitrary $\ell $. We have done this only for the simplest cases $%
\ell =1,2$ and $3$. Let's consider these cases separately.

\subsection{Case $\ell =1$}

In this case, $\underline{s}=$\underline{$\sigma $}$\left( \tau \right) $,
where $\tau $ is the duration of the NCP and \underline{$\sigma $}$\left(
\tau \right) $ is given in Eqs. (\ref{54}). In Eqs. (\ref{45}) Raman
frequency detuning consists of two terms, $\nu ^{\left( 0\right) }$ and $\nu
^{\left( 1\right) }$. The contribution to the $s-$matrix (\ref{54b}) from
the term $\nu ^{\left( 1\right) }$ can be estimated as 
\end{subequations}
\begin{equation}
\delta \underline{s}=\underline{s}^{\prime }\nu ^{\left( 1\right) }\sim
\left( \mathbf{k}\cdot \mathbf{g}-\alpha \right) T\tau ,  \label{63}
\end{equation}%
where%
\begin{equation}
\underline{s}^{\prime }\equiv \partial \underline{s}/\partial \nu ,
\label{64}
\end{equation}%
and we take into account that the characteristic size of the $s-$matrix
dependence on $\nu $, is of the order of\textbf{\ }$\mathbf{\tau }^{-1}$.
The parameter $\left( \mathbf{k}\cdot \mathbf{g}-\alpha \right) T\tau $ is
responsible for the corrections to the MZAI phase caused by the finite
duration of the Raman pulses \cite{i29,i31,i31.1}. We consider this
parameter to be small,%
\begin{equation}
\delta \phi \sim \left\vert \mathbf{k}\cdot \mathbf{g}-\alpha \right\vert
T\tau \ll 1,  \label{65}
\end{equation}%
and calculate the MZAI phase up to a correction linear in $\tau $.

Here and below we reserve the denotations $\nu $ and $\nu _{r}$ for the
Raman frequency detuning on the nonresonant and resonant branches of the
recoil diagram, respectively. The effective Rabi frequency%
\begin{equation}
\left\vert \Omega \right\vert =\dfrac{\pi }{\tau }  \label{66}
\end{equation}%
Let us first consider the nonresonant branch of the recoil diagram. In the
zero approximation in $\delta \phi $ we are looking for such a duration $%
\tau $ of the auxiliary Raman pulse, at which the atom does not change its
state with a probability of $100\%$,%
\begin{equation}
\left\vert s_{gg}\right\vert =1.  \label{67}
\end{equation}%
From Eq. (\ref{54c}), the solution to this equation is%
\begin{equation}
\Omega _{r}\tau =2j\pi ,  \label{68}
\end{equation}%
where $j$ is an arbitrary positive integer. From Eqs. (\ref{66}, \ref{68})
one finds that%
\begin{equation}
\tau =\pi \dfrac{\sqrt{4j^{2}-1}}{\left\vert \nu ^{\left( 0\right)
}\right\vert }.  \label{69}
\end{equation}%
In this case, the $s-$matrix is%
\begin{equation}
\underline{\sigma }\left( \tau \right) =\left( -1\right) ^{j}I,  \label{69,1}
\end{equation}%
where $I$ is the identity matrix. Consider now the amendments (\ref{63}).
Since 
\begin{subequations}
\label{70}
\begin{eqnarray}
s_{d}^{\prime } &=&-\dfrac{\nu \tau }{2\Omega _{r}}\sin \dfrac{\Omega
_{r}\tau }{2}+i\left\{ \dfrac{1}{\Omega _{r}}\sin \dfrac{\Omega _{r}\tau }{2}%
+\dfrac{\nu ^{2}\tau }{2\Omega _{r}^{2}}\cos \dfrac{\Omega _{r}\tau }{2}-%
\dfrac{\nu ^{2}}{\Omega _{r}^{3}}\sin \dfrac{\Omega _{r}\tau }{2}\right\} ,
\label{70a} \\
s_{a}^{\prime } &=&\dfrac{\Omega \nu }{\Omega _{r}}\left( -\dfrac{1}{\Omega
_{r}^{2}}\sin \dfrac{\Omega _{r}\tau }{2}+\dfrac{\tau }{2\Omega _{r}}\cos 
\dfrac{\Omega _{r}\tau }{2}\right) .  \label{70b}
\end{eqnarray}%
Then from Eqs. (\ref{66}, \ref{68}), one gets that, taking into account
corrections linear in $\tau $, the $s-$matrix is 
\end{subequations}
\begin{subequations}
\label{71}
\begin{eqnarray}
\underline{s} &=&\left( 
\begin{array}{cc}
\exp \left\{ i\left[ \eta _{0}^{\left( 1\right) }-\eta _{1}^{\left( 1\right)
}\left( \nu \right) \nu ^{\left( 1\right) }\right] \right\} & -i\left(
-1\right) ^{j}\left[ \pi \Omega \left( 4j^{2}-1\right) /8j^{2}\left\vert
\Omega \right\vert \nu \right] \nu ^{\left( 1\right) } \\ 
-i\left( -1\right) ^{j}\left[ \pi \Omega ^{\ast }\left( 4j^{2}-1\right)
/8j^{2}\left\vert \Omega \right\vert \nu \right] \nu ^{\left( 1\right) } & 
\exp \left\{ i\left[ \eta _{0}^{\left( 1\right) }+\eta _{1}^{\left( 1\right)
}\left( \nu \right) \nu ^{\left( 1\right) }\right] \right\}%
\end{array}%
\right) ,  \label{71a} \\
\eta _{0}^{\left( 1\right) } &=&j\pi ,  \label{71b} \\
\eta _{1}^{\left( 1\right) }\left( \nu \right) &=&-\dfrac{\pi \left(
4j^{2}-1\right) ^{3/2}}{8j^{2}\left\vert \nu \right\vert }.  \label{71c}
\end{eqnarray}%
One sees that the correction (\ref{45c}) due to the small but nonzero pulse
duration results in small off-diagonal $s-$matrix elements. This means that
even an ideal rectangular pulse leads to a small splitting of the atomic
state on the nonresonant branch of the recoil diagram. In the following, we
will only take into account the influence of corrections (\ref{45c}) on the
AI phase, and therefore we will neglect the off-diagonal elements of the $s-$%
matrix (\ref{71a}).

Let us now turn to the resonance branch. Here, the $\pi -$pulse must, with $%
100\%$ probability, transfer the atom from one internal state to another,
while transmitting the momentum $\pm \hbar \mathbf{k}$. However, at $\nu =0,$
from Eqs. (\ref{70}), derivatives $\left\{ s_{d}^{\prime },s_{a}^{\prime
}\right\} =\left\{ i\dfrac{\tau }{\pi },0\right\} ,$ so that 
\end{subequations}
\begin{equation}
\underline{s}=\left( 
\begin{array}{cc}
i\tau \nu _{r}^{\left( 1\right) }/\pi & -i\exp \left( i\arg \Omega \right)
\\ 
-i\exp \left( -i\arg \Omega \right) & -i\tau \nu _{r}^{\left( 1\right) }/\pi%
\end{array}%
\right) .  \label{72}
\end{equation}%
One sees that the mirror, the resonant $\pi -$pulse, ceases to be ideal
owing to the finite duration of the pulse. With a small amplitude linear in
this duration, the momentum of the atom and its internal state remain
unchanged. The diagonal elements of the $s-$matrix (\ref{72}) do not affect
the phase of the MZAI, and we neglect them below.

\subsection{Case $\ell =2$}

We have already noted that the disadvantage of a single Raman pulse is that
for each new value of the Raman detuning $\nu ^{\left( 0\right) }$ one must
change the pulse duration according to Eq. (\ref{69}) and then change the
Rabi frequency according to Eq. (\ref{66}). For $\ell >1$ we will consider
NCPs, in which the Rabi frequency is the same as that of the main pulses,
and for given durations of rectangular pulses $\tau _{m}$ $\left( m=1,\ldots
,\ell \right) $, the condition \ref{II} is reached owing to properly chosen
time delays between pulses $\tau _{bm},$ $\left( m=1,\ldots ,\ell -1\right) $%
. Here and below, one reserves the denotation $\tau $ only for the duration
of the first main $\dfrac{\pi }{2}-$pulse, so for all pulses the magnitude
of the Rabi frequency is%
\begin{equation}
\left\vert \Omega \right\vert =\dfrac{\pi }{2\tau }  \label{72.1}
\end{equation}

\subsubsection{\textbf{Resonant branch, }$\mathbf{\protect\nu }_{r}^{\left( 
\mathbf{0}\right) }\mathbf{=0}$}

For the $s-$matrix (\ref{55}) one obtains%
\begin{equation}
\underline{s}_{bm}=I+\dfrac{i}{2}\nu ^{\left( 1\right) }\tau _{bm}\underline{%
\sigma }_{3},  \label{73}
\end{equation}%
where \underline{$\sigma $}$_{3}$ is the Pauli matrix. Then 
\begin{subequations}
\label{74}
\begin{eqnarray}
\underline{s} &=&\underline{\sigma }\left( \dsum_{m=1}^{\ell }\tau
_{m}\right) +\delta \underline{s}_{b},  \label{74a} \\
\delta \underline{s}_{b} &=&\dfrac{i}{2}\nu ^{\left( 1\right)
}\dsum_{m=1}^{\ell -1}\tau _{bm}\underline{\sigma }\left( \dsum_{m^{\prime
}=m+1}^{\ell }\tau _{m^{\prime }}\right) \underline{\sigma }_{3}\underline{%
\sigma }\left( \dsum_{m^{\prime }=1}^{m}\tau _{m^{\prime }}\right) .
\label{74b}
\end{eqnarray}%
Here we have used the law of multiplication for $s-$matrices, 
\end{subequations}
\begin{equation}
\underline{\sigma }\left( \tau _{m}\right) \underline{\sigma }\left( \tau
_{m-1}\right) =\underline{\sigma }\left( \tau _{m}+\tau _{m-1}\right)
\label{75}
\end{equation}%
In order for the NCP to be a $\pi -$pulse, it is necessary that the sum of
durations $\tau _{m}$ be equal to $2\tau $,%
\begin{equation}
\dsum_{m=1}^{\ell }\tau _{m}=2\tau .  \label{76}
\end{equation}%
Then, after the change $\tau \rightarrow 2\tau $ in Eq. (\ref{72}), one
obtains%
\begin{equation}
\underline{\sigma }\left( 2\tau \right) =\left( 
\begin{array}{cc}
2i\tau \nu _{r}^{\left( 1\right) }/\pi & -i\exp \left( i\arg \Omega \right)
\\ 
-i\exp \left( -i\arg \Omega \right) & -2i\tau \nu _{r}^{\left( 1\right) }/\pi%
\end{array}%
\right) .  \label{77}
\end{equation}%
For $\ell =2$ we have considered only the symmetric case, when two
rectangular pulses have the same duration%
\begin{equation}
\tau _{1}=\tau _{2}=\tau ,  \label{78}
\end{equation}%
when $\delta \underline{s}_{b}=\dfrac{i}{2}\nu ^{\left( 1\right) }\tau _{b}$%
\underline{$\sigma $}$_{3},$ with $\tau _{b}\equiv \tau _{b1},$ and
therefore 
\begin{equation}
\underline{s}=\left( 
\begin{array}{cc}
i\left( 2\tau /\pi +\tau _{b}/2\right) \nu _{r}^{\left( 1\right) } & -i\exp
\left( i\arg \Omega \right) \\ 
-i\exp \left( -i\arg \Omega \right) & -i\left( 2\tau /\pi +\tau
_{b}/2\right) \nu _{r}^{\left( 1\right) }%
\end{array}%
\right) .  \label{79}
\end{equation}

\subsubsection{\textbf{Nonresonant branch.}}

In this case one will get for $s-$matrix (\ref{57})%
\begin{equation}
\underline{s}=\underline{\sigma }\left( \tau \right) \underline{s}_{b1}%
\underline{\sigma }\left( \tau \right) ,  \label{79.1}
\end{equation}%
where $\underline{\sigma }\left( \tau \right) $ and $\underline{s}_{b1}$ are
given by Eqs. (\ref{54}, \ref{55}). Multiplying the matrices one gets%
\begin{equation}
\underline{s}=\left( 
\begin{array}{cc}
\exp \left[ \dfrac{i}{2}\nu \tau _{b}\right] s_{d}^{2}-\exp \left[ -\dfrac{i%
}{2}\nu \tau _{b}\right] \left\vert s_{a}\right\vert ^{2} & -i\exp \left[ 
\dfrac{i}{2}\nu \tau _{b}\right] s_{a}s_{d}-i\exp \left[ -\dfrac{i}{2}\nu
\tau _{b}\right] s_{a}s_{d}^{\ast } \\ 
-i\exp \left[ \dfrac{i}{2}\nu \tau _{b}\right] s_{a}^{\ast }s_{d}-i\exp %
\left[ -\dfrac{i}{2}\nu \tau _{b}\right] s_{a}^{\ast }s_{d}^{\ast } & \exp 
\left[ -\dfrac{i}{2}\nu \tau _{b}\right] s_{d}^{\ast 2}-\exp \left[ \dfrac{i%
}{2}\nu \tau _{b}\right] \left\vert s_{a}\right\vert ^{2}%
\end{array}%
\right) .  \label{80}
\end{equation}%
One sees that for equal durations $\tau _{1}$ and $\tau _{2}$%
\begin{subequations}
\label{80.1}
\begin{eqnarray}
s_{ee} &=&s_{gg}^{\ast },  \label{80.1a} \\
s_{ge} &=&\exp \left( -2i\arg \Omega \right) s_{eg.}  \label{80.1b}
\end{eqnarray}%
Condition \ref{II} is satisfied if the delay between pulses, $\tau _{b}$, is
the root of the equation 
\end{subequations}
\begin{equation}
s_{eg}=0.  \label{81}
\end{equation}%
Subject to Eq. (\ref{72.1}), one gets%
\begin{equation}
\tau _{b}=\tau ^{\left( 2\right) }\left( j,\nu \right) =\dfrac{1}{\nu }\left[
-2\arctan \left( \dfrac{2\nu \tau }{\sqrt{\pi ^{2}+4\nu ^{2}\tau ^{2}}}\tan 
\dfrac{1}{4}\sqrt{\pi ^{2}+4\nu ^{2}\tau ^{2}}\right) +\func{sgn}\left( \nu
\right) \left( 2j+1\right) \pi \right] .  \label{82}
\end{equation}%
Since $\tau _{b}>0$, then $j$ must be a non-negative integer.

In contrast to the SLMT in the Bragg regime \cite{i10}, in our case,
although the NCP does not excite the atom and does not change its momentum,
it leads to the appearance of a phase factor in the atomic wave function.
From Eqs. (\ref{80}, \ref{82}) one will get for this factor 
\begin{subequations}
\label{83}
\begin{eqnarray}
s_{gg} &=&\exp \left\{ i\left[ \eta _{0}^{\left( 2\right) }+\eta
_{0}^{\left( 2\right) }\left( \nu \right) \right] \right\} ,  \label{83a} \\
\eta _{0}^{\left( 2\right) } &=&\left[ j-\dfrac{1}{2}\func{sgn}\left( \nu
\right) \right] \pi ,  \label{83b} \\
\eta _{0}^{\left( 2\right) }\left( \nu \right) &=&-\arctan \left[ \dfrac{%
2\nu \tau }{\sqrt{\pi ^{2}+4\nu ^{2}\tau ^{2}}}\tan \left( \dfrac{1}{4}\sqrt{%
\pi ^{2}+4\nu ^{2}\tau ^{2}}\right) \right] .  \label{83c}
\end{eqnarray}%
Let us now turn to the calculation of the correction (\ref{63}). Direct
calculation of the derivative (\ref{64}) turned out to be unproductive.
Note, however, that the $s-$matrix is a function of two variables $\nu $ and 
$\tau _{b}$ and therefore 
\end{subequations}
\begin{equation}
\dfrac{d}{d\nu }\left. \underline{s}\right\vert _{\tau _{b}=\tau ^{\left(
2\right) }\left( j,\nu \right) }=\underline{s}^{\prime }+\left. \dfrac{%
\partial \underline{s}}{\partial \tau _{b}}\right\vert _{\tau _{b}=\tau
^{\left( 2\right) }\left( j,\nu \right) }\dfrac{d\tau ^{\left( 2\right)
}\left( j,\nu \right) }{d\nu }.  \label{84}
\end{equation}%
From Eqs. (\ref{83a}, \ref{82}, \ref{80}) it follows, correspondingly, that 
\begin{subequations}
\label{85}
\begin{eqnarray}
\dfrac{d}{d\nu }\left. s_{gg}\right\vert _{\tau _{b}=\tau ^{\left( 2\right)
}\left( j,\nu \right) } &=&-is_{gg}\func{Im}\dfrac{s_{d}^{\prime }}{s_{d}},
\label{85a} \\
\dfrac{d\tau ^{\left( 2\right) }\left( j,\nu \right) }{d\nu } &=&-\dfrac{1}{%
\nu }\tau ^{\left( 2\right) }\left( j,\nu \right) -\dfrac{2}{\nu }\func{Im}%
\left( \dfrac{s_{d}^{\prime }}{s_{d}}\right) ,  \label{85b} \\
\left. \dfrac{\partial s_{gg}}{\partial \tau _{b}}\right\vert _{\tau
_{b}=\tau ^{\left( 2\right) }\left( j,\nu \right) } &=&\dfrac{i}{2}\nu
s_{gg}\left( \left\vert s_{a}\right\vert ^{2}-\left\vert s_{d}\right\vert
^{2}\right) ,  \label{85c}
\end{eqnarray}%
and, substituting these values into Eq. (\ref{84}), one calculates
consequently $s_{gg}^{\prime }$ and $\delta s_{gg}$. The correction $\delta
s_{eg}$ is calculated in a similar way. As a result, taking into account
Eqs. (\ref{80.1}) one arrives at the following expression for the $s-$%
matrix: 
\end{subequations}
\begin{subequations}
\label{86}
\begin{gather}
\underline{s}=\left( 
\begin{array}{cc}
\exp \left\{ -i\left[ \eta _{0}^{\left( 2\right) }+\eta _{0}^{\left(
2\right) }\left( \nu ^{\left( 0\right) }\right) +\eta _{1}^{\left( 2\right)
}\left( \nu ^{\left( 0\right) }\right) \nu ^{\left( 1\right) }\right]
\right\} & -\left( -1\right) ^{j}\func{sgn}\left( \nu ^{\left( 0\right)
}\right) s_{a}\left\vert s_{d}\right\vert \left[ \tau _{b}+2\func{Im}\left( 
\dfrac{s_{d}^{\prime }}{s_{d}}\right) \right] \nu ^{\left( 1\right) } \\ 
-\left( -1\right) ^{j}\func{sgn}\left( \nu ^{\left( 0\right) }\right)
s_{a}^{\ast }\left\vert s_{d}\right\vert \left[ \tau _{b}+2\func{Im}\left( 
\dfrac{s_{d}^{\prime }}{s_{d}}\right) \right] \nu ^{\left( 1\right) } & \exp
\left\{ i\left[ \eta _{0}^{\left( 2\right) }+\eta _{0}^{\left( 2\right)
}\left( \nu ^{\left( 0\right) }\right) +\eta _{1}^{\left( 2\right) }\left(
\nu ^{\left( 0\right) }\right) \nu ^{\left( 1\right) }\right] \right\}%
\end{array}%
\right) ,  \label{86a} \\
\eta _{1}^{\left( 2\right) }\left( \nu \right) =-\dfrac{\tau ^{\left(
2\right) }\left( j,\nu \right) }{2}\left( \left\vert s_{d}\right\vert
^{2}-\left\vert s_{a}\right\vert ^{2}\right) -2\left\vert s_{d}\right\vert
^{2}\func{Im}\dfrac{s_{d}^{\prime }}{s_{d}}.  \label{86b}
\end{gather}%
As in the preceding section, we will further neglect the diagonal elements
of the resonant $s-$matrix (\ref{79}) and the off-diagonal elements of the
nonresonant $s-$matrix (\ref{86a}).

\subsection{Case $\ell =3$}

Here we also consider only the symmetric case, when 
\end{subequations}
\begin{subequations}
\label{87}
\begin{eqnarray}
\tau _{3} &=&\tau _{1},  \label{87a} \\
\tau _{2} &=&2\left( \tau -\tau _{1}\right) ,  \label{87b} \\
\tau _{b1} &=&\tau _{b2}\equiv \tau _{b}.  \label{87c}
\end{eqnarray}

\subsubsection{\textbf{Resonant branch, }$\protect\nu _{r}^{\left( 0\right) }%
\mathbf{=0}$.}

Here the $\sigma -$matrix is given in Eq. (\ref{77}). Then, computing the
matrix (\ref{74b}) using Eqs. (\ref{54c}, \ref{54d}, \ref{87}) one arrives
at the following expression for the resonant $s-$matrix: 
\end{subequations}
\begin{equation}
\underline{s}=\left( 
\begin{array}{cc}
i\nu _{r}^{\left( 1\right) }\left\{ \tau _{b}\cos \left[ \dfrac{\pi }{2}%
\left( 1-\dfrac{\tau _{1}}{\tau }\right) \right] +2\dfrac{\tau }{\pi }%
\right\} & -i\exp \left( i\arg \Omega \right) \\ 
-i\exp \left( -i\arg \Omega \right) & -i\nu _{r}^{\left( 1\right) }\left\{
\tau _{b}\cos \left[ \dfrac{\pi }{2}\left( 1-\dfrac{\tau _{1}}{\tau }\right) %
\right] +2\dfrac{\tau }{\pi }\right\}%
\end{array}%
\right) .  \label{88}
\end{equation}

\subsubsection{\textbf{Nonresonant branch.}}

\begin{figure}[!t]
\includegraphics[width=12cm]{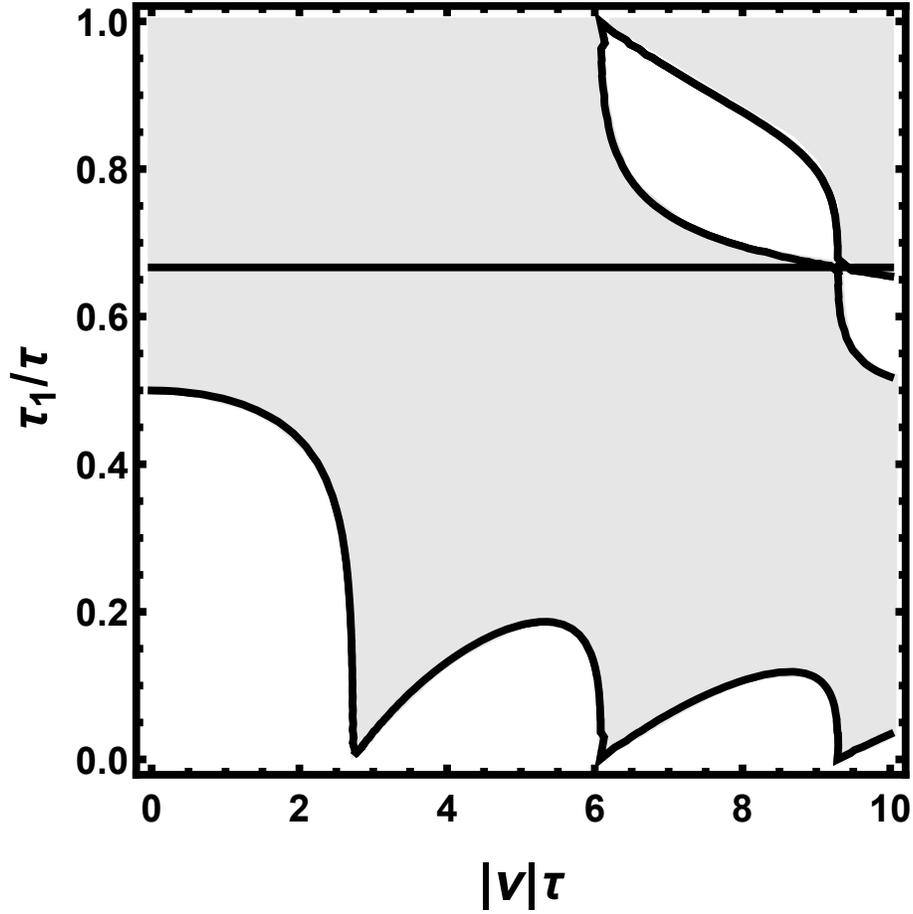}
\caption{For a given value of the Raman detuning $\protect\nu $, the
duration of the first and third rectangular pulses should be in the gray
area. The straight line corresponds to pulses of equal duration, $\protect%
\tau _{1}=\protect\tau _{2}=\protect\tau _{3}=\dfrac{2}{3}\protect\tau ,$
when the solution of the Eq. (\protect\ref{90}) exists for any $\protect\nu $%
.}
\label{f4}
\end{figure}

In this case, the $s-$matrix (\ref{57}) is given by%
\begin{equation}
\underline{s}=\underline{\sigma }\left( \tau _{1}\right) \underline{s}_{b}%
\underline{\sigma }\left( \tau _{2}\right) \underline{s}_{b}\underline{%
\sigma }\left( \tau _{1}\right)  \label{88.1}
\end{equation}%
where $\underline{\sigma }\left( \tau \right) $ and $\underline{s}_{b}$ are
given by Eqs. (\ref{54}, \ref{55}). Taking into account the fact that the
effective Rabi frequency $\Omega $ does not change for all components of the
symmetric NCP of rectangular pulses and multiplying the matrices in Eq. (\ref%
{88.1}) one arrives at the result

\begin{gather}
\underline{s}=  \notag \\
\left( 
\begin{array}{cc}
\exp \left( i\nu \tau _{b}\right) s_{d1}^{2}s_{d2}-\exp \left( -i\nu \tau
_{b}\right) \left\vert s_{a1}\right\vert ^{2}s_{d2}^{\ast
}-2s_{d1}\left\vert s_{a2}s_{a1}\right\vert & -i\left\{ 2s_{a1}\func{Re}%
\left[ \exp \left( i\nu \tau _{b}\right) s_{d1}s_{d2}\right] +s_{a2}\left(
\left\vert s_{d1}\right\vert ^{2}-\left\vert s_{a1}\right\vert ^{2}\right)
\right\} \\ 
-i\left\{ s_{a1}^{\ast }2\func{Re}\left[ \exp \left( i\nu \tau _{b}\right)
s_{d2}s_{d1}\right] +s_{a2}^{\ast }\left( \left\vert s_{d1}\right\vert
^{2}-\left\vert s_{a1}\right\vert ^{2}\right) \right\} & \exp \left( -i\nu
\tau _{b}\right) s_{d1}^{\ast 2}s_{d2}^{\ast }-\exp \left( i\nu \tau
_{b}\right) s_{d2}\left\vert s_{a1}\right\vert ^{2}-2s_{d1}^{\ast
}\left\vert s_{a2}s_{a1}\right\vert%
\end{array}%
\right) ,  \label{89}
\end{gather}%
where 
\begin{subequations}
\label{89.1}
\begin{eqnarray}
s_{di} &=&s_{d}\left( \tau _{i}\right) ,  \label{89.1a} \\
s_{ai} &=&s_{a}\left( \tau _{i}\right)  \label{89.1b}
\end{eqnarray}%
If one chooses the delay between rectangular pulses, $\tau _{b}$, in such a
way that 
\end{subequations}
\begin{equation}
\cos \left[ \nu \tau _{b}+\arg \left( s_{d1}s_{d2}\right) \right] =\dfrac{%
\left\vert s_{a2}\right\vert \left( \left\vert s_{a1}\right\vert
^{2}-\left\vert s_{d1}\right\vert ^{2}\right) }{2\left\vert
s_{a1}s_{d1}s_{d2}\right\vert },  \label{90}
\end{equation}%
then $s_{eg}=0$ and therefore the NCP satisfies requirement \ref{II}. Figure %
\ref{f4} shows (in gray) the range for which Eq. (\ref{90}) has a solution.

The solution of the Eq. (\ref{90}) is given by 
\begin{subequations}
\label{91}
\begin{eqnarray}
\tau _{b} &=&\dfrac{1}{2}\tau ^{\left( 3_{\pm }\right) }\left( j,\nu \right)
,  \label{91a} \\
\tau ^{\left( 3_{\pm }\right) }\left( j,\nu \right) &=&\dfrac{2}{\nu }\left[
\pm \arccos \dfrac{\left\vert s_{a2}\right\vert \left( \left\vert
s_{a1}\right\vert ^{2}-\left\vert s_{d1}\right\vert ^{2}\right) }{%
2\left\vert s_{a1}s_{d1}s_{d2}\right\vert }-\arctan \dfrac{\func{Im}\left(
s_{d1}s_{d2}\right) }{\func{Re}\left( s_{d1}s_{d2}\right) }+2j\pi \right] ,
\label{91b}
\end{eqnarray}%
where for each value of $\nu $ the integer $j$ can take only those values
for which $\tau _{b}>0$. Hence, the wave function of the atom in the ground
state, after the action of the NCP acquires the phase factor 
\end{subequations}
\begin{subequations}
\label{92}
\begin{eqnarray}
s_{gg} &=&\exp \left[ i\eta _{0}^{\left( 3_{\pm }\right) }+i\eta
_{0}^{\left( 3_{\pm }\right) }\left( \nu \right) \right] ,  \label{92a} \\
\eta _{0}^{\left( 3_{\pm }\right) } &=&\pi ,  \label{92b} \\
\eta _{0}^{\left( 3_{\pm }\right) }\left( \nu \right) &=&-\arctan \left( 
\dfrac{2\tau \nu }{\sqrt{\pi ^{2}+4\nu ^{2}\tau ^{2}}}\tan \dfrac{\tau _{1}}{%
4\tau }\sqrt{\pi ^{2}+4\nu ^{2}\tau ^{2}}\right) \pm \arccos \dfrac{1}{2}%
\left\vert \dfrac{s_{a2}}{s_{a1}s_{d1}}\right\vert .  \label{92c}
\end{eqnarray}%
To calculate linear in $\nu ^{\left( 1\right) }$ corrections, one can use
the equality (\ref{84}), in which one should make the substitution 
\end{subequations}
\begin{equation}
\tau ^{\left( 2\right) }\left( j,\nu \right) \rightarrow \dfrac{1}{2}\tau
^{\left( 3_{\pm }\right) }\left( j,\nu \right) .  \label{92.1}
\end{equation}%
From Eqs. (\ref{92a}, \ref{89}) follows, respectively, that 
\begin{subequations}
\label{93}
\begin{eqnarray}
\dfrac{d}{d\nu }\left. s_{gg}\right\vert _{\tau _{b}=0.5\tau ^{\left( 3_{\pm
}\right) }\left( j,\nu \right) } &=&i\dfrac{d\eta _{0}^{\left( 3_{\pm
}\right) }\left( \nu \right) }{d\nu }\left. s_{gg}\right\vert _{\tau
_{b}=0.5\tau ^{\left( 3_{\pm }\right) }\left( j,\nu \right) },  \label{93a}
\\
\left. \dfrac{\partial s_{gg}}{\partial \tau _{b}}\right\vert _{\tau
_{b}=0.5\tau ^{\left( 3_{\pm }\right) }\left( j,\nu \right) } &=&i\nu \left(
\left\vert s_{a1}\right\vert ^{2}-\left\vert s_{d1}\right\vert ^{2}\right)
\left. s_{gg}\right\vert _{\tau _{b}=0.5\tau ^{\left( 3_{\pm }\right)
}\left( j,\nu \right) }.  \label{93b}
\end{eqnarray}%
These expressions allow one to extract $s_{gg}^{\prime }$ from Eq. (\ref{84}%
) [taking into account the replacement (\ref{92.1})] and get $\delta s_{gg}$%
. Computing the off-diagonal element $s_{eg},$ in a similar way, one arrives
at the following result 
\end{subequations}
\begin{subequations}
\label{94}
\begin{gather}
\underline{s}=  \notag \\
\left( 
\begin{array}{cc}
\exp \left[ -i\eta _{0}^{\left( 3_{\pm }\right) }-i\eta _{0}^{\left( 3_{\pm
}\right) }\left( \nu \right) -i\eta _{1}^{\left( 3_{\pm }\right) }\left( \nu
\right) \nu _{1}\right] & \mp i\exp \left( i\arg \Omega \right) \dfrac{d\tau
^{\left( 3_{\pm }\right) }}{d\nu }\sqrt{4\left\vert s_{a1}\right\vert
^{2}\left\vert s_{d1}\right\vert ^{2}-\left\vert s_{a2}\right\vert ^{2}}\nu
^{\left( 0\right) }\nu ^{\left( 1\right) } \\ 
\mp i\exp \left( -i\arg \Omega \right) \dfrac{d\tau ^{\left( 3_{\pm }\right)
}}{d\nu }\sqrt{4\left\vert s_{a1}\right\vert ^{2}\left\vert
s_{d1}\right\vert ^{2}-\left\vert s_{a2}\right\vert ^{2}}\nu ^{\left(
0\right) }\nu ^{\left( 1\right) } & \exp \left[ i\eta _{0}^{\left( 3_{\pm
}\right) }+i\eta _{0}^{\left( 3_{\pm }\right) }\left( \nu \right) +i\eta
_{1}^{\left( 3_{\pm }\right) }\left( \nu \right) \nu _{1}\right]%
\end{array}%
\right) ,  \label{94a} \\
\eta _{1}^{\left( 3_{\pm }\right) }\left( \nu \right) =\dfrac{d\eta
_{0}^{\left( 3_{\pm }\right) }\left( \nu \right) }{d\nu }-\dfrac{\nu }{2}%
\left( \left\vert s_{a1}\right\vert ^{2}-\left\vert s_{d1}\right\vert
^{2}\right) \dfrac{d\tau ^{\left( 3_{\pm }\right) }\left( j,\nu \right) }{%
d\nu }.  \label{94b}
\end{gather}%
From Eqs. (\ref{56}, \ref{69}) the total NCP duration is given by 
\end{subequations}
\begin{subequations}
\label{94.1}
\begin{eqnarray}
\tau \left( \nu ,\zeta ,\beta ,1\right) &=&\pi \dfrac{\sqrt{4j_{\zeta ,\beta
,1}^{2}-1}}{\left\vert \nu \right\vert }  \label{94.1a} \\
\tau \left( \nu ,\zeta ,\beta ,2\right) &=&2\tau +\tau ^{\left( 2\right)
}\left( j_{\zeta ,\beta ,2},\nu \right) ,  \label{94.1b} \\
\tau \left( \nu ,\zeta ,\beta ,3_{\pm }\right) &=&2\tau +\tau ^{\left(
3_{\pm }\right) }\left( j_{\zeta ,\beta ,3_{\pm }},\nu \right) ,
\label{94.1c}
\end{eqnarray}%
where $\tau ^{\left( 2\right) }\left( j,\nu \right) $ and $\tau ^{\left(
3_{\pm }\right) }\left( j,\nu \right) $ are given in Eqs. (\ref{82}, \ref%
{91b}) respectively.

\section{\label{sIII}The phase}

Consider the interaction of an atomic cloud with a sequence of three main
resonant Raman pulses $\pi /2-\pi -\pi /2$, acting at the moments $\left\{
T_{1,0},T_{2,0},T_{3,0}\right\} $, having duration $\tau -2\tau -\tau $ and
effective wave vector $\mathbf{k}$. Suppose that the atom was launched at $%
t=0$ in the ground state, i.e. at $t<T_{1,0}$%
\end{subequations}
\begin{subequations}
\label{95}
\begin{eqnarray}
c\left( e,\mathbf{p},t\right) &=&0,  \label{95a} \\
c\left( g,\mathbf{p},t\right) &=&\sqrt{f_{g}\left( \mathbf{p}\right) },
\label{95b}
\end{eqnarray}%
where $f_{g}\left( \mathbf{p}\right) $ is the momentum distribution function
in the atomic cloud. Below, up to Eq. (\ref{113}), to simplify the
calculation, we omit the factor $\sqrt{f_{g}\left( \mathbf{p}\right) }$.
Regarding the auxiliary Raman $\pi -$pulses, we will assume that there are
an even number of them, $2n$, in each of the four sets. All pulses, main and
auxiliary, are numbered with integers $\left\{ \zeta ,\beta \right\} $,
where $1\leq \zeta \leq 3$. The numbers of the main pulses are $\left\{
\zeta ,0\right\} $, and for the auxiliary NCPs at $\zeta =1$ the values $%
1\leq \beta \leq 2n$ correspond to the set following the first main $\pi /2-$%
pulse, at $\zeta =2$ the values $-2n\leq \beta \leq -1$ and $1\leq \beta
\leq 2n$ correspond to the pulses that preceded or followed after the second
main $\pi -$pulse, with $\zeta =3$ the values $-2n\leq \beta \leq -1$
correspond to the set preceding the third main $\pi /2-$pulse. Each of the
pulses is NCP, consisting of $1\leq \ell _{\zeta ,\beta }\leq 3$ rectangular
pulses. The effective wave vector and momentum chirp rate are 
\end{subequations}
\begin{subequations}
\begin{eqnarray}
\mathbf{k}_{\zeta ,\beta } &=&\left( -1\right) ^{\beta }\mathbf{k},
\label{96a} \\
\alpha _{\zeta ,\beta } &=&\left( -1\right) ^{\beta }\alpha  \label{96b}
\end{eqnarray}%
and its total duration $\tau \left( \nu ,\zeta ,\beta ,\ell _{\zeta ,\beta
}\right) $ for auxiliary NCPs is given in Eqs. (\ref{94.1}), and for the
main ones there are 
\end{subequations}
\begin{equation}
\tau \left( 0,1,0,1\right) =\tau \left( 0,2,0,1\right) /2=\tau \left(
0,3,0,1\right) =\tau ,  \label{96.1}
\end{equation}%
here one takes into account that the main pulses are resonant for the blue
and red branches of the recoil diagram, i.e. for them $\nu ^{\left( 0\right)
}=0$. In our calculations, we take into account the finite durations of the
Raman pulses. For their timing, following Ref. \cite{i29}, we introduce a
delay time between pulses%
\begin{equation}
d_{\zeta ,m}\equiv \left\{ 
\begin{array}{c}
T_{\zeta ,m}-T_{\zeta ,m-1}-\tau \left( \nu ,\zeta ,m-1,\ell _{\zeta
,m-1}\right) ,\text{ for }-2n<m\leq 2n \\ 
T\equiv T_{\zeta ,-2n}-T_{\zeta -1,2n}-\tau \left( \nu ,\zeta -1,2n,\ell
_{\zeta -1,2n}\right) ,\text{ for }m=-2n,\zeta >1%
\end{array}%
\right. ,  \label{96.2}
\end{equation}%
i.e., we define the interrogation time $T$ as the time between the end of
the last auxiliary NCP $\left\{ 1,2n\right\} $ and the beginning of the
first auxiliary NCP $\left\{ 2,-2n\right\} $, which coincides with the time
between the end of the last auxiliary NCP $\left\{ 2,2n\right\} $ and the
beginning of the first auxiliary NCP $\left\{ 3,-2n\right\} $. From the
given $T_{1,0}$, delays between NCPs (\ref{96.2}) and their durations (\ref%
{94.1}, \ref{96.1}) one can make up the full timing , the moments of action
of each of the pulses $T_{\zeta ,\beta }$. One can represent them as%
\begin{equation}
T_{\zeta ,\beta }=T_{\zeta ,\beta }^{\left( 0\right) }+\Upsilon _{\zeta
,\beta },  \label{96.3}
\end{equation}%
where $T_{\zeta ,\beta }^{\left( 0\right) }$ is $T_{1,0}$ plus the sum of
all preceding delays, while $\Upsilon _{\zeta ,\beta }$ is the sum of the
durations of all preceding pulses.

In this article, we neglect the deviations of pulses from the ideal shape
and duration calculated below. Therefore, the gray lines on the recoil
diagram associated with these deviations do not appear. We also assume
negligible diagonal elements in $s-$matrices (\ref{72}, \ref{79}, \ref{88})
and off-diagonal elements in $s-$matrices (\ref{71a}, \ref{86a}, \ref{94a}).
So the gray lines associated with the finite duration of the Raman pulses do
not appear either. As a result, only two branches, blue and red, remain on
the recoil diagram. In this case, the wave function of an atom is a column,%
\begin{equation}
\underline{\chi }=\left( 
\begin{array}{c}
\chi _{b} \\ 
\chi _{r}%
\end{array}%
\right) .  \label{97}
\end{equation}

After the action of $\pi /2-$pulse $\left\{ 1,0\right\} $ from Eqs. (\ref{59}%
, \ref{95}) one will find that this column is%
\begin{equation}
\left( 
\begin{array}{c}
c\left( e,\mathbf{p}_{+}^{\left( 1,0\right) }+\hbar \mathbf{k},T_{1,0}+\tau
\right) \\ 
c\left( g,\mathbf{p}_{+}^{\left( 1,0\right) },T_{1,0}+\tau \right)%
\end{array}%
\right) =\left( 
\begin{array}{c}
S_{eg}^{\left( 1,0\right) }\left( \mathbf{p}_{+,T_{1,0}+\tau }^{\left(
1,0\right) }+\hbar \mathbf{k}/2\right) \\ 
S_{gg}^{\left( 1,0\right) }\left( \mathbf{p}_{+,T_{1,0}+\tau }^{\left(
1,0\right) }+\hbar \mathbf{k}/2\right)%
\end{array}%
\right) ,  \label{98}
\end{equation}%
where the matrix $S\left( \mathbf{P}\right) $ is given in Eq. (\ref{60a}).
At resonance, when%
\begin{equation}
\nu _{r}^{\left( 0\right) }=\delta _{1,0}^{\left( 0\right) }\left( \mathbf{p}%
_{+,T_{1,0}+\tau }^{\left( 1,0\right) }+\hbar \mathbf{k}/2\right) =\delta
_{1,0}-\dfrac{\mathbf{k}}{M}\mathbf{\cdot p}_{+,T_{1,0}+\tau }^{\left(
1,0\right) }-\omega _{k}\approx 0  \label{99}
\end{equation}%
and $\left\vert \Omega _{1,0}\right\vert $ given in Eq. (\ref{72.1}), from
Eqs. (\ref{54b}, \ref{45a}, \ref{45d}, \ref{70a}) it follows, respectively,
that

\begin{subequations}
\label{100}
\begin{eqnarray}
\underline{s} &=&\dfrac{1}{\sqrt{2}}\left( 
\begin{array}{cc}
1 & -i\exp \left( i\arg \Omega _{1,0}\right) \\ 
-i\exp \left( i\arg \Omega _{1,0}\right) & 1%
\end{array}%
\right) ,  \label{100a} \\
\nu _{r}^{\left( 0\right) } &=&\delta _{1,0}-\dfrac{\mathbf{k}}{M}\cdot 
\mathbf{p}_{+,T_{1,0}+\tau }^{\left( 1,0\right) }-\omega _{k}  \label{100b}
\\
\nu _{r} &=&-\left( \mathbf{k\cdot g}-\alpha \right) T_{1,0},  \label{100c}
\\
\phi _{1,0}\left( \mathbf{p}_{+,T_{1,0}+\tau }^{\left( 1,0\right) }+\hbar 
\mathbf{k}/2\right) &=&\phi _{1,0}+\left( \delta _{1,0}-\dfrac{\mathbf{k}}{M}%
\mathbf{\cdot p}_{+,T_{1,0}+\tau }^{\left( 1,0\right) }-\omega _{k}\right)
T_{1,0}-\dfrac{1}{2}\left( \mathbf{k\cdot g}-\alpha \right) T_{1,0}^{2},
\label{100d} \\
s_{d}^{\prime } &=&\dfrac{i\sqrt{2}}{\pi },  \label{100e}
\end{eqnarray}%
Despite the resonance, we retained the Ramsey term \cite{i28} [the second
term in the expression for the phase (\ref{100d})]. For resonance it is
enough that 
\end{subequations}
\begin{equation}
\left\vert \nu _{r}^{\left( 0\right) }\right\vert \ll \tau ^{-1}.
\label{100.1}
\end{equation}%
If%
\begin{equation}
T_{1,0}\sim T\gg \tau ,  \label{100.2}
\end{equation}%
then, despite the resonance condition, one can observe Ramsey fringes at%
\begin{equation}
T^{-1}\lesssim \left\vert \nu _{r}^{\left( 0\right) }\right\vert \ll \tau
^{-1}.  \label{100.3}
\end{equation}%
From Eqs.. (\ref{98}, \ref{60}, \ref{100}) one gets 
\begin{subequations}
\label{101}
\begin{gather}
\left( 
\begin{array}{c}
c\left( e,\mathbf{p}_{+}^{\left( 1,0\right) }+\hbar \mathbf{k},T_{1,0}+\tau
\right) \\ 
c\left( g,\mathbf{p}_{+}^{\left( 1,0\right) },T_{1,0}+\tau \right)%
\end{array}%
\right) =\dfrac{1}{\sqrt{2}}\left( 
\begin{array}{c}
\exp \left( i\psi _{1,0}^{\left( b\right) }\right) \\ 
\exp \left( i\psi _{1,0}^{\left( r\right) }\right)%
\end{array}%
\right) ,  \label{101a} \\
\psi _{1,0}^{\left( b\right) }=-\dfrac{\pi }{2}-\phi _{1,0}+\arg \Omega
_{1,0}-\left( \delta _{1,0}-\dfrac{\mathbf{k}}{M}\cdot \mathbf{p}%
_{+,T_{1,0}+\tau }^{\left( 1,0\right) }-\omega _{k}\right) T_{1,0}+\dfrac{1}{%
2}\left( \mathbf{k\cdot g}-\alpha \right) T_{1,0}^{2}+\dfrac{1}{2}\left( 
\mathbf{k\cdot g}-\alpha \right) T_{1,0}\tau ,  \label{101b} \\
\psi _{1,0}^{\left( r\right) }=\left( \dfrac{2}{\pi }-\dfrac{1}{2}\right)
\left( \mathbf{k\cdot g}-\alpha \right) T_{1,0}\tau .  \label{101c}
\end{gather}%
%
With the exception of the last beam splitter $\left\{ 3,0\right\} $, all
subsequent beam splitters are $\pi -$pulses. An ideal $\pi -$pulse does not
change the magnitude of states, only the phases of these states change.
These changes, phase augmentations, in sum determine the phase of the
interferometer.%
%

Let us now consider the action of an odd NCP $\left\{ 1,2m-1\right\} $, for
which $\mathbf{k}_{1,2m-1}=-\mathbf{k}.$ Before interaction 
\end{subequations}
\begin{equation}
\underline{\chi }=\left( 
\begin{array}{c}
c\left( e,\mathbf{\pi }_{-},T_{1,2m-1}\right) \\ 
c\left( g,\mathbf{p}_{-}^{\left( 1,2m-1\right) },T_{1,2m-1}\right)%
\end{array}%
\right) =\dfrac{1}{\sqrt{2}}\left( 
\begin{array}{c}
\exp \left( i\psi _{1,2m-2}^{\left( b\right) }\right) \\ 
\exp \left( i\psi _{1,2m-2}^{\left( r\right) }\right)%
\end{array}%
\right) .  \label{102}
\end{equation}%
From the recoil diagrams in Figs. \ref{f2}, \ref{f3}, one can conclude that
on the resonant branch, for arbitrary $m$, the preceding Raman pulses
transferred to the atom an odd number of momenta $\hbar \mathbf{k}$, i.e.%
\begin{equation}
\mathbf{\pi }_{-}=\mathbf{p}_{-}^{\left( 1,2m-1\right) }+\left( 2m-1\right)
\hbar \mathbf{k}  \label{102.1}
\end{equation}%
Then, along the blue resonance branch after the NCP, from Eqs. (\ref{60.2b}, %
\ref{96a}) for $\mathfrak{N}=2m$ the state of the atom changes as 
\begin{subequations}
\label{103}
\begin{eqnarray}
c\left( g,\mathbf{\pi }_{+},T_{1,2m-1}+\tau _{p}\right) &=&S_{ge}^{\left(
1,2m-1\right) }\left( \mathbf{\pi }_{s}\right) c\left( e,\mathbf{\pi }%
_{-},T_{1,2m-1}\right)  \label{103a} \\
\mathbf{\pi }_{+} &=&\mathbf{p}_{+}^{\left( 1,2m-1\right) }+2m\hbar \mathbf{k%
},  \label{103b} \\
\tau _{p} &=&\tau \left( \nu ,1,2m-1,\ell _{1,2m-1}\right)  \label{103c} \\
\mathbf{\pi }_{s} &=&\mathbf{\pi +}\left( 4m-1\right) \hbar \mathbf{k}/2,
\label{103d} \\
\mathbf{\pi } &=&\mathbf{p}_{+,T_{1,2m-1}+\tau _{p}}^{\left( 1,2m-1\right) }
\label{103e}
\end{eqnarray}%
In this case, under the condition of resonance 
\end{subequations}
\begin{subequations}
\label{104}
\begin{gather}
\nu _{r}^{\left( 0\right) }=\delta _{1,2m-1}^{\left( 0\right) }\left( 
\mathbf{\pi }_{s}\right) =\left\{ \delta _{1,2m-1}+\dfrac{\mathbf{k}}{M}%
\mathbf{\cdot \pi }\right\} +\left( 4m-1\right) \omega _{k}\approx 0,
\label{104a} \\
\nu _{r}=\left( \mathbf{k\cdot g}-\alpha \right) T_{1,2m-1},  \label{104b} \\
\phi _{1,2m-1}\left( \mathbf{\pi }_{s}\right) =\phi _{1,2m-1}+\left[ \delta
_{1,2m-1}+\dfrac{\mathbf{k}}{M}\mathbf{\cdot \pi }+\left( 4m-1\right) \omega
_{k}\right] T_{1,2m-1}+\dfrac{1}{2}\left( \mathbf{k\cdot g}-\alpha \right)
T_{1,2m-1}^{2},  \label{104c}
\end{gather}%
and then, using the off-diagonal matrix element in the $s-$matrices (\ref{72}%
, \ref{77}, \ref{88}), from Eqs.. (\ref{103}, \ref{60}, \ref{104}) one
receives 
\end{subequations}
\begin{subequations}
\label{105}
\begin{gather}
c\left( g,\mathbf{\pi }_{+},T_{1,2m-1}+\tau _{p}\right) =\dfrac{1}{\sqrt{2}}%
\exp \left( i\psi _{1,2m-1}^{\left( b\right) }\right) ,  \label{105a} \\
\psi _{1,2m-1}^{\left( b\right) }=\psi _{1,2m-2}^{\left( b\right)
}+A_{1,2m-1}^{\left( b\right) },  \label{105b} \\
A_{1,2m-1}^{\left( b\right) }=-\dfrac{\pi }{2}+\phi _{1,2m-1}-\arg \Omega
_{1,2m-1}+\left[ \delta _{1,2m-1}+\dfrac{\mathbf{k}}{M}\mathbf{\cdot \pi }%
+\left( 4m-1\right) \omega _{k}\right] T_{1,2m-1}^{\left( 0\right) }  \notag
\\
+\dfrac{1}{2}\left( \mathbf{k\cdot g}-\alpha \right) T_{1,2m-1}^{2}+\dfrac{1%
}{2}\left( \mathbf{k\cdot g}-\alpha \right) T_{1,2m-1}^{\left( 0\right)
}\tau _{p},  \label{105c}
\end{gather}%
where $A_{1,2m-1}^{\left( b\right) }$ is the phase augmentation of the
atomic amplitude on the blue branch during interaction with the NCP $\left\{
1,2m-1\right\} $. In the expression (\ref{105c}) we have replaced 
\end{subequations}
\begin{equation}
T_{\zeta ,\beta }\rightarrow T_{\zeta ,\beta }^{\left( 0\right) }
\label{105.1}
\end{equation}%
in the 4th and last term. This is because in the resonance condition (\ref%
{100.1})%
\begin{equation}
\left\vert \nu _{r}\right\vert \Upsilon _{\zeta ,\beta }\ll 1.  \label{105.2}
\end{equation}%
We also took into account that one can neglect the term $\Upsilon _{\zeta
,\beta }\tau $ since it is bilinear in pulse durations.

Consider now the nonresonant red branch, where%
\begin{equation}
c\left( g,\mathbf{p}_{+}^{\left( 1,2m-1\right) },T_{1,2m-1}+\tau _{p}\right)
=S_{gg}^{\left( 1,2m-1\right) }\left( \mathbf{\pi }-\hbar \mathbf{k}%
/2\right) c\left( g,\mathbf{p}_{-}^{\left( 1,2m-1\right) },T_{1,2m-1}\right)
.  \label{106}
\end{equation}%
From Eqs. (\ref{45}, \ref{104a}) one finds%
\begin{equation}
\nu =\delta _{1,2m-1}\left( \mathbf{\pi }-\hbar \mathbf{k}/2\right) =\left\{
\delta _{1,2m-1}+\dfrac{\mathbf{k}}{M}\mathbf{\cdot \pi }\right\} -\omega
_{k}+\left( \mathbf{k\cdot g}-\alpha \right) T_{1,2m-1}.  \label{107}
\end{equation}%
Since terms in curly brackets in Eqs. (\ref{104a}, \ref{107}) coincide, then%
\begin{equation}
\nu =-4m\omega _{k}+\left( \mathbf{k\cdot g}-\alpha \right) T_{1,2m-1}.
\label{108}
\end{equation}%
%
Thus, knowing that there is no frequency detuning on the resonant branch of
the recoil diagram, one is able to determine the detuning $\nu $ on the
nonresonant branch. Since at a negligibly small recoil frequency there is no
detuning on both branches, then at $\omega _{k}\neq 0$ the detuning $\nu $
is determined only by the recoil frequency and the Doppler frequency shift
does not contribute to it.%
%

The second term in Eq.(\ref{108}) is a small correction associated with the
finite duration of the NCP. One sees that in order to fulfill the
requirement \ref{II}, the total duration of the NCP must be equal to 
\begin{equation}
\tau _{p}=\tau \left( -4m\omega _{k},1,2m-1,\ell _{1,2m-1}\right)
\label{108.1}
\end{equation}%
given by Eqs. (\ref{94.1}). Then using the matrix element $s_{gg}$ $s-$%
matrices (\ref{71a}, \ref{86a}, \ref{94a}), from Eqs. (\ref{106}, \ref{60}, %
\ref{108}) one finds that 
\begin{subequations}
\label{109}
\begin{gather}
c\left( g,\mathbf{p}_{+}^{\left( 1,2m-1\right) },T_{1,2m-1}+\tau _{p}\right)
=\dfrac{1}{\sqrt{2}}\exp \left( i\psi _{1,2m-1}^{\left( r\right) }\right) ,
\label{109a} \\
\psi _{1,2m-1}^{\left( r\right) }=\psi _{1,2m-2}^{\left( r\right)
}+A_{1,2m-1}^{\left( r\right) },  \label{109b} \\
A_{1,2m-1}^{\left( r\right) }=-2m\omega _{k}\tau _{p}+\eta _{0}^{\left( \ell
_{1,2m-1}\right) }+\eta _{0}^{\left( \ell _{1,2m-1}\right) }\left( -4m\omega
_{k}\right) +\left[ \dfrac{1}{2}\tau _{p}+\eta _{1}^{\left( \ell
_{1,2m-1}\right) }\left( -4m\omega _{k}\right) \right] \left( \mathbf{k\cdot
g}-\alpha \right) T_{1,2m-1}^{\left( 0\right) },  \label{109c}
\end{gather}%
where $\eta _{0}^{\left( \ell _{1,2m-1}\right) }$ is given by Eqs. (\ref{71b}%
, \ref{83b}, \ref{92b}), 
\end{subequations}
\begin{equation}
\eta _{0}^{\left( 1\right) }\left( \nu \right) =0,  \label{110}
\end{equation}%
$\eta _{0}^{\left( 2\right) }\left( \nu \right) $ and $\eta _{0}^{\left(
3_{\pm }\right) }\left( \nu \right) $ are respectively given by Eqs. (\ref%
{83c}, \ref{92c}\TEXTsymbol{>}), $\eta _{1}^{\left( \ell \right) }\left( \nu
\right) $ are given by Eqs. (\ref{71c}, \ref{86b}, \ref{94b}). In the last
term, we also made a replacement (\ref{105.1}).

The action of the remaining $\pi -$pulses can be calculated in a similar
way. One can verify that the ideal $\pi -$pulse $\left\{ \zeta ,\beta
\right\} $, for which one assigns the duration exactly according to the
expressions (\ref{94.1}), and which is in resonance with atomic transitions
with an accuracy much less than the inverse pulse duration, leads only to
phase augmentation of the atomic states amplitudes, $A_{\zeta ,\beta
}^{\left( b,r\right) }$. One arrives at the following expressions for these
augmentations 
\begin{subequations}
\label{111}
\begin{gather}
A_{1,m}^{\left( b\right) }=-\dfrac{\pi }{2}-\left( -1\right) ^{m}\left( \phi
_{1,m}-\arg \Omega _{1,m}\right) -\left[ \left( -1\right) ^{m}\delta _{1,m}-%
\dfrac{\mathbf{k}}{M}\mathbf{\cdot \pi }-\left( 2m+1\right) \omega _{k}%
\right] T_{1,m}^{\left( 0\right) }  \notag \\
+\dfrac{1}{2}\left( \mathbf{k\cdot g}-\alpha \right) T_{1,m}^{2}+\dfrac{1}{2}%
\left( \mathbf{k\cdot g}-\alpha \right) T_{1,m}^{\left( 0\right) }\tau _{p},
\label{111a} \\
A_{1,m}^{\left( r\right) }=\left( -1\right) ^{m}\xi _{m}\omega _{k}\tau
_{p}+\eta _{0}^{\left( \ell _{1,m}\right) }+\eta _{0}^{\left( \ell
_{1,m}\right) }\left[ \left( -1\right) ^{m}2\xi _{m}\omega _{k}\right]
-\left( -1\right) ^{m}\left\{ \dfrac{1}{2}\tau _{p}+\eta _{1}^{\left( \ell
_{1,m}\right) }\left[ \left( -1\right) ^{m}2\xi _{m}\omega _{k}\right]
\right\} \left( \mathbf{k\cdot g}-\alpha \right) T_{1,m}^{\left( 0\right) },
\label{111b} \\
A_{2,-m}^{\left( b\right) }=-\dfrac{\pi }{2}+\left( -1\right) ^{m}\left(
\phi _{2,-m}-\arg \Omega _{2,-m}\right) +\left[ \left( -1\right) ^{m}\delta
_{2,-m}-\dfrac{\mathbf{k}}{M}\mathbf{\cdot \pi }-\left( 2m+1\right) \omega
_{k}\right] T_{2,-m}^{\left( 0\right) }  \notag \\
-\dfrac{1}{2}\left( \mathbf{k\cdot g}-\alpha \right) T_{2,-m}^{2}-\dfrac{1}{2%
}\left( \mathbf{k\cdot g}-\alpha \right) T_{2,-m}^{\left( 0\right) }\tau
_{p},  \label{111c} \\
A_{2,-m}^{\left( r\right) }=\left( -1\right) ^{m}\xi _{m}\omega _{k}\tau
_{p}+\eta _{0}^{\left( \ell _{2,-m}\right) }+\eta _{0}^{\left( \ell
_{2,-m}\right) }\left( \left( -1\right) ^{m}2\xi _{m}\omega _{k}\right)
-\left( -1\right) ^{m}\left[ \dfrac{1}{2}\tau _{p}+\eta _{1}^{\left( \ell
_{2,-m}\right) }\left( \left( -1\right) ^{m}2\xi _{m}\omega _{k}\right) %
\right] \left( \mathbf{k\cdot g}-\alpha \right) T_{2,-m}^{\left( 0\right) },
\label{111d} \\
A_{2,0}^{\left( b\right) }=-\dfrac{\pi }{2}+\phi _{2,0}-\arg \Omega _{2,0}+%
\left[ \delta _{2,0}-\dfrac{\mathbf{k}}{M}\mathbf{\cdot p}_{+,T_{2,0}+2\tau
}^{\left( 2,0\right) }-\omega _{k}\right] T_{2,0}^{\left( 0\right) }-\dfrac{1%
}{2}\left( \mathbf{k\cdot g}-\alpha \right) T_{2,0}^{2}-\left( \mathbf{%
k\cdot g}-\alpha \right) T_{2,0}^{\left( 0\right) }\tau ,  \label{111e} \\
A_{2,0}^{\left( r\right) }=-\pi -A_{2,0}^{\left( b\right) },  \label{111f} \\
A_{2,m}^{\left( b\right) }=\left( -1\right) ^{m}\xi _{m}\omega _{k}\tau
_{p}+\eta _{0}^{\left( \ell _{2,m}\right) }+\eta _{0}^{\left( \ell
_{2,m}\right) }\left( \left( -1\right) ^{m}2\xi _{m}\omega _{k}\right)
-\left( -1\right) ^{m}\left[ \dfrac{1}{2}\tau _{p}+\eta _{1}^{\left( \ell
_{2,m}\right) }\left( \left( -1\right) ^{m}2\xi _{m}\omega _{k}\right) %
\right] \left( \mathbf{k\cdot g}-\alpha \right) T_{2,m}^{\left( 0\right) },
\label{111g} \\
A_{2,m}^{\left( r\right) }=-\dfrac{\pi }{2}-\left( -1\right) ^{m}\left( \phi
_{2,m}-\arg \Omega _{2,m}\right) -\left[ \left( -1\right) ^{m}\delta _{2,m}-%
\dfrac{\mathbf{k}}{M}\mathbf{\cdot \pi }-\left( 2m+1\right) \omega _{k}%
\right] T_{2,m}^{\left( 0\right) }  \notag \\
+\dfrac{1}{2}\left( \mathbf{k\cdot g}-\alpha \right) T_{2,m}^{2}+\dfrac{1}{2}%
\left( \mathbf{k\cdot g}-\alpha \right) T_{2,m}^{\left( 0\right) }\tau _{p},
\label{111h} \\
A_{3,-m}^{\left( b\right) }=\left( -1\right) ^{m}\xi _{m}\omega _{k}\tau
_{p}+\eta _{0}^{\left( \ell _{3,-m}\right) }+\eta _{0}^{\left( \ell
_{3,-m}\right) }\left( \left( -1\right) ^{m}2\xi _{m}\omega _{k}\right)
-\left( -1\right) ^{m}\left[ \dfrac{1}{2}\tau _{p}+\eta _{1}^{\left( \ell
_{3,-m}\right) }\left( \left( -1\right) ^{m}2\xi _{m}\omega _{k}\right) %
\right] \left( \mathbf{k\cdot g}-\alpha \right) T_{3,-m}^{\left( 0\right) },
\label{111i} \\
A_{3,-m}^{\left( r\right) }=-\dfrac{\pi }{2}+\left( -1\right) ^{m}\left(
\phi _{3,-m}-\arg \Omega _{3,-m}\right) +\left[ \left( -1\right) ^{m}\delta
_{3,-m}-\dfrac{\mathbf{k}}{M}\mathbf{\cdot \pi }-\left( 2m+1\right) \omega
_{k}\right] T_{3,-m}^{\left( 0\right) }  \notag \\
-\dfrac{1}{2}\left( \mathbf{k\cdot g}-\alpha \right) T_{3,-m}^{2}-\dfrac{1}{2%
}\left( \mathbf{k\cdot g}-\alpha \right) T_{3,-m}^{\left( 0\right) }\tau
_{p},  \label{111j}
\end{gather}%
where 
\end{subequations}
\begin{equation}
\xi _{m}=m+\dfrac{1}{2}\left( 1-\left( -1\right) ^{m}\right) .  \label{112}
\end{equation}%
In the expressions for augmentations $A_{\zeta ,\beta }^{\left( b,r\right) }$
the duration of the pulse $\left\{ \zeta ,\beta \right\} $ $\tau _{p}$ and
momentum $\mathbf{\pi }$ are given by 
\begin{subequations}
\label{112.1}
\begin{eqnarray}
\tau _{p} &=&\tau \left[ \left( -1\right) ^{\beta }2\xi _{\beta }\omega
_{k},\zeta ,\beta ,\ell _{\zeta ,\beta }\right] ,  \label{112.1a} \\
\mathbf{\pi } &\mathbf{=}&\mathbf{p}_{+,T_{\zeta ,\beta }+\tau _{p}}^{\left(
\zeta ,\beta \right) }  \label{112.1b}
\end{eqnarray}%
%
Augmentations (\ref{111a}, \ref{111c}, \ref{111e}, \ref{111f}, \ref{111h}, %
\ref{111j}) refer to the resonant branch of the recoil diagram, while
augmentations (\ref{111b}, \ref{111d}, \ref{111g}, \ref{111i}) refer to the
nonresonant branch.%
%
The factor $(-1)^{m}$ reflects the fact that at the transitions $%
g\rightarrow e$ and $e\rightarrow g$ one uses different offdiagonal elements
of the matrix $S$ in Eq. (\ref{60}), and for them the signs of the phase
factors are opposite. As a result, the terms associated with the detuning $%
\delta $ change sign. But at the same time, the terms associated with the
Doppler shift and the gravitational field remain unchanged, because for
transitions $g\rightarrow e$ and $e\rightarrow g$ one uses opposite
effective wave vectors $\pm \mathbf{k.}$

Using augmentations (\ref{111}), one calculates the phases of the atomic
states amplitudes before the action of the third main $\pi /2-$pulse, $%
\left\{ \psi _{3,-1}^{\left( b\right) },\psi _{3,-1}^{\left( r\right)
}\right\} $. We are interested in the total probability of excitation of
atoms in the cloud 
\end{subequations}
\begin{equation}
w=\int d\mathbf{p}_{+}^{\left( 3,0\right) }f_{g}\left[ \mathbf{p}%
_{+}^{\left( 3,0\right) }\right] \left\vert c\left( e,\mathbf{p}_{+}^{\left(
3,0\right) }+\hbar \mathbf{k},T_{3,0}+\tau \right) \right\vert ^{2},
\label{113}
\end{equation}%
where we have restored the factor $\sqrt{f_{g}\left( \mathbf{p}\right) }$
from Eq. (\ref{95b}). At $t=T_{3,0}+\tau $ the amplitude of an atom in an
excited state consists of blue and red components,%
\begin{equation}
c\left( e,\mathbf{p}_{+}^{\left( 3,0\right) }+\hbar \mathbf{k},T_{3,0}+\tau
\right) =c^{\left( b\right) }\left( e,\mathbf{p}_{+}^{\left( 3,0\right)
}+\hbar \mathbf{k},T_{3,0}+\tau \right) +c^{\left( r\right) }\left( e,%
\mathbf{p}_{+}^{\left( 3,0\right) }+\hbar \mathbf{k},T_{3,0}+\tau \right) ,
\label{114}
\end{equation}%
for which one has 
\begin{subequations}
\label{115}
\begin{eqnarray}
c^{\left( b\right) }\left( e,\mathbf{p}_{+}^{\left( 3,0\right) }+\hbar 
\mathbf{k},T_{3,0}+\tau \right) &=&S_{eg}^{\left( 3,0\right) }\left( \mathbf{%
p}_{+,T_{3,0}+\tau }^{\left( 3,0\right) }+\hbar \mathbf{k}/2\right) c\left(
g,\mathbf{p}_{-}^{\left( 3,0\right) },T_{3,0}\right) ,  \label{115a} \\
c^{\left( r\right) }\left( e,\mathbf{p}_{+}^{\left( 3,0\right) }+\hbar 
\mathbf{k},T_{3,0}+\tau \right) &=&S_{ee}^{\left( 3,0\right) }\left( \mathbf{%
p}_{+,T_{3,0}+\tau }^{\left( 3,0\right) }+\hbar \mathbf{k}/2\right) c\left(
e,\mathbf{p}_{-}^{\left( 3,0\right) }+\hbar \mathbf{k},T_{3,0}\right) .
\label{115b}
\end{eqnarray}%
After calculations similar to those used in the derivation of Eqs. (\ref{101}%
), one obtains 
\end{subequations}
\begin{subequations}
\label{116}
\begin{gather}
\left( 
\begin{array}{c}
c^{\left( b\right) }\left( e,\mathbf{p}_{+}^{\left( 3,0\right) }+\hbar 
\mathbf{k},T_{3,0}+\tau \right) \\ 
c^{\left( r\right) }\left( e,\mathbf{p}_{+}^{\left( 3,0\right) }+\hbar 
\mathbf{k},T_{3,0}+\tau \right)%
\end{array}%
\right) =\dfrac{1}{2}\left( 
\begin{array}{c}
\exp \left( i\psi _{3,0}^{\left( b\right) }\right) \\ 
\exp \left( i\psi _{3,0}^{\left( r\right) }\right)%
\end{array}%
\right) ;  \label{116a} \\
\psi _{3,0}^{\left( b,r\right) }=\psi _{3,-1}^{\left( b,r\right)
}+A_{3,0}^{\left( b,r\right) },  \label{116b} \\
A_{3,0}^{\left( b\right) }=-\dfrac{\pi }{2}-\phi _{3,0}+\arg \Omega _{3,0}-%
\left[ \delta _{3,0}-\dfrac{\mathbf{k}}{M}\mathbf{\cdot p}_{+,T_{3,0}+\tau
}^{\left( 3,0\right) }-\omega _{k}\right] T_{3,0}^{\left( 0\right) }+\dfrac{1%
}{2}\left( \mathbf{k\cdot g}-\alpha \right) T_{3,0}^{2}+\dfrac{1}{2}\left( 
\mathbf{k\cdot g}-\alpha \right) T_{3,0}\tau ,  \label{116c} \\
A_{3,0}^{\left( r\right) }=\left( \dfrac{1}{2}-\dfrac{2}{\pi }\right) \left( 
\mathbf{k\cdot g}-\alpha \right) T_{3,0}^{\left( 0\right) }\tau .
\label{116d}
\end{gather}

In Eqs. (\ref{114}) the independent variable is the momentum of the atom
after the action of the pulse $\left\{ 3,0\right\} $, $\mathbf{p}%
_{+}^{\left( 3,0\right) }$. Using Eqs. (\ref{61}, \ref{62b}) one can
calculate the values of all other momenta in Eqs. (\ref{111}). Such
calculations would be necessary if we were only interested in one of the
ports on Figs. \ref{f1}-\ref{f3}. The total excitation probability (\ref{113}%
) is the sum of the probabilities over all possible ports. And for this
response, in Eq. (\ref{113}), one introduces a new integration variable 
\end{subequations}
\begin{equation}
\mathbf{p}_{i}=\mathbf{p}_{+,T_{3,0}+\tau }^{\left( 3,0\right) }.
\label{117}
\end{equation}%
From Eq. (\ref{58b}) it follows that all pulses in Eqs. (\ref{111}) coincide
with $\mathbf{p}_{i}$.%
%
This coincidence is a consequence of the fact that in Eq. (\ref{60.1}) one
selected in the momentum the parts associated with the transfer of photon
momenta $\mathfrak{N\hbar }\mathbf{k}$, and the momentum that changes only
under the action of the gravitational field.%
%

Integrand in Eq. (\ref{113}) is a rapidly oscillating function momentum $%
\mathbf{p}_{i}$ with a period of the order of $M/kT$. One chooses the timing
of the pulses in the AI such that these oscillations disappear. If, in
addition, the atomic cloud is cooled to such a temperature that the momentum
in the functions of the parameters $s_{d}$ and $s_{a}$ from Eqs. (\ref{54})
can be considered to be the same as the average momentum in the cloud, then
one will receive%
\begin{equation}
w=\dfrac{1}{2}\left( 1-\cos \phi \right) ,  \label{118}
\end{equation}%
where the phase of the atomic interferometer is 
\begin{subequations}
\label{119}
\begin{eqnarray}
\phi &=&\pi +\psi _{3,0}^{\left( b\right) }-\psi _{3,0}^{\left( r\right)
}=\pi +\dsum_{j=1}^{3}A_{j,0}+\dsum_{m=1}^{2n}\left(
A_{1,m}+A_{2,-m}+A_{2,m}+A_{3,-m}\right) ,  \label{119a} \\
A_{\zeta ,\beta } &\equiv &\left\{ 
\begin{array}{c}
\psi _{1,0}^{\left( b\right) }-\psi _{1,0}^{\left( r\right) },\text{ for }%
\left\{ \zeta ,\beta \right\} =\left\{ 1,0\right\} \\ 
A_{\zeta ,\beta }^{\left( b\right) }-A_{\zeta ,\beta }^{\left( r\right) },%
\text{ for }\left\{ \zeta ,\beta \right\} \not=\left\{ 1,0\right\}%
\end{array}%
\right. .  \label{119b}
\end{eqnarray}%
Then from Eqs. (\ref{101b}, \ref{101c}, \ref{111}, \ref{116c}, \ref{116d})
one arrives at the next result 
\end{subequations}
\begin{equation}
\phi =\bar{\phi}+\phi _{R}+\phi _{D}+\phi _{q}+\bar{\phi}_{q}+\phi
_{g}^{\left( 0\right) }+\phi _{g}^{\left( 1\right) },  \label{120}
\end{equation}%
where%
\begin{equation}
\bar{\phi}=-\phi _{1,0}+2\phi _{2,0}-\phi _{3,0}-\dsum_{m=1}^{2n}\left\{
\left( -1\right) ^{m}\left[ \phi _{1,m}-\phi _{2,-m}-\phi _{2,m}+\phi _{3,-m}%
\right] +\eta _{0}^{\left( \ell _{1,m}\right) }+\eta _{0}^{\left( \ell
_{2,-m}\right) }-\eta _{0}^{\left( \ell _{2,m}\right) }-\eta _{0}^{\left(
\ell _{3,-m}\right) }\right\} ,  \label{121}
\end{equation}%
Ramsey phase%
\begin{equation}
\phi _{R}=-\delta _{1,0}T_{1,0}+2\delta _{2,0}T_{2,0}^{\left( 0\right)
}-\delta _{3,0}T_{3,0}^{\left( 0\right) }+\dsum_{m=1}^{2n}\left( -1\right)
^{m}\left( -\delta _{1,m}T_{1,m}^{\left( 0\right) }+\delta
_{2,-m}T_{2,-m}^{\left( 0\right) }+\delta _{2,m}T_{2,m}^{\left( 0\right)
}-\delta _{3,-m}T_{3,-m}^{\left( 0\right) }\right) ,  \label{122}
\end{equation}%
Doppler phase%
\begin{equation}
\phi _{D}=\dfrac{\mathbf{k}}{M}\mathbf{\cdot p}_{i}\left[ T_{1,0}-2T_{2,0}^{%
\left( 0\right) }+T_{3,0}^{\left( 0\right) }+\dsum_{m=1}^{2n}\left(
T_{1,m}^{\left( 0\right) }-T_{2,-m}^{\left( 0\right) }-T_{2,m}^{\left(
0\right) }+T_{3,-m}^{\left( 0\right) }\right) \right] ,  \label{123}
\end{equation}%
quantum phases 
\begin{subequations}
\label{124}
\begin{gather}
\phi _{q}=\omega _{k}\left[ T_{1,0}-2T_{2,0}^{\left( 0\right)
}+T_{3,0}^{\left( 0\right) }+\dsum_{m=1}^{2n}\left( 2m+1\right) \left(
T_{1,m}^{\left( 0\right) }-T_{2,-m}^{\left( 0\right) }-T_{2,m}^{\left(
0\right) }+T_{3,-m}^{\left( 0\right) }\right) \right] ,  \label{124a} \\
\bar{\phi}_{q}=\dsum_{m=1}^{2n}\{-\eta _{0}^{\left( \ell _{1,m}\right) }%
\left[ \left( -1\right) ^{m}2\xi _{m}\omega _{k}\right] -\eta _{0}^{\left(
\ell _{2,-m}\right) }\left[ \left( -1\right) ^{m}2\xi _{m}\omega _{k}\right]
+\eta _{0}^{\left( \ell _{2,m}\right) }\left[ 2\left( -1\right) ^{m}\xi
_{m}\omega _{k}\right]  \notag \\
+\eta _{0}^{\left( \ell _{3,-m}\right) }\left( \left( -1\right) ^{m}2\xi
_{m}\omega _{k}\right) +\left( -1\right) ^{m}\xi _{m}\omega _{k}[-\tau
\left( \left( -1\right) ^{m}2\xi _{m}\omega _{k},1,m,\ell _{1,m}\right)
-\tau \left( \left( -1\right) ^{m}2\xi _{m}\omega _{k},2,-m,\ell
_{2,-m}\right)  \notag \\
+\tau \left( \left( -1\right) ^{m}2\xi _{m}\omega _{k},2,m,\ell
_{2,m}\right) +\tau \left( \left( -1\right) ^{m}2\xi _{m}\omega
_{k},3,-m,\ell _{3,-m}\right) ]\},  \label{124b}
\end{gather}%
gravity phases 
\end{subequations}
\begin{subequations}
\label{125}
\begin{gather}
\phi _{g}^{\left( 0\right) }=\dfrac{1}{2}\left( \mathbf{k\cdot g}-\alpha
\right) \left[ T_{1,0}^{\left( 0\right) 2}-2T_{2,0}^{\left( 0\right)
2}+T_{3,0}^{\left( 0\right) 2}+\dsum_{m=1}^{2n}\left( T_{1,m}^{\left(
0\right) 2}-T_{2,-m}^{\left( 0\right) 2}-T_{2,m}^{\left( 0\right)
2}+T_{3,-m}^{\left( 0\right) 2}\right) \right] ,  \label{125a} \\
\phi _{g}^{\left( 1\right) }=\left( \mathbf{k\cdot g}-\alpha \right) \left\{
T_{1,0}^{\left( 0\right) }\Upsilon _{1,0}-2T_{2,0}^{\left( 0\right)
}\Upsilon _{2,0}+T_{3,0}^{\left( 0\right) }\Upsilon _{3,0}+\tau \left[
\left( 1-\dfrac{2}{\pi }\right) T_{1,0}^{\left( 0\right) }-2T_{2,0}^{\left(
0\right) }+\dfrac{2}{\pi }T_{3,0}^{\left( 0\right) }\right] \right.  \notag
\\
+\dsum_{m=1}^{2n}\left[ T_{1,m}^{\left( 0\right) }\Upsilon
_{1,m}-T_{2,-m}^{\left( 0\right) }\Upsilon _{2,-m}-T_{2,m}^{\left( 0\right)
}\Upsilon _{2,m}+T_{3,-m}^{\left( 0\right) }\Upsilon _{3,-m}\right.  \notag
\\
+\left( \varsigma _{m+1}\tau \left( \left( -1\right) ^{m}2\xi _{m}\omega
_{k},1,m,\ell _{1,m}\right) +\left( -1\right) ^{m}\eta _{1}^{\left( \ell
_{1,m}\right) }\left( \left( -1\right) ^{m}2\xi _{m}\omega _{k}\right)
\right) T_{1,m}^{\left( 0\right) }  \notag \\
-\left( \varsigma _{m}\tau \left( \left( -1\right) ^{m}2\xi _{m}\omega
_{k},2,-m,\ell _{2,-m}\right) -\left( -1\right) ^{m}\eta _{1}^{\left( \ell
_{2,-m}\right) }\left( \left( -1\right) ^{m}2\xi _{m}\omega _{k}\right)
\right) T_{2,-m}^{\left( 0\right) }  \notag \\
-\left( \varsigma _{m+1}\tau \left( \left( -1\right) ^{m}2\xi _{m}\omega
_{k},2,m,\ell _{2,m}\right) +\left( -1\right) ^{m}\eta _{1}^{\left( \ell
_{2,m}\right) }\left( 2\left( -1\right) ^{m}\xi _{m}\omega _{k}\right)
\right) T_{2,m}^{\left( 0\right) }  \notag \\
\left. \left. +\left( \varsigma _{m}\tau \left( \left( -1\right) ^{m}2\xi
_{m}\omega _{k},3,-m,\ell _{3,-m}\right) -\left( -1\right) ^{m}\eta
_{1}^{\left( \ell _{3,-m}\right) }\left( 2\left( -1\right) ^{m}\xi
_{m}\omega _{k}\right) \right) T_{3,-m}^{\left( 0\right) }\right] \right\} ,
\label{125b}
\end{gather}%
and parameter 
\end{subequations}
\begin{equation}
\varsigma _{m}\equiv \dfrac{1}{2}\left( 1-\left( -1\right) ^{m}\right) .
\label{126}
\end{equation}%
\twocolumngrid%

\section{\label{sIV}Doppler and quantum phases}

\begin{figure}[t]
\includegraphics[width=8cm]{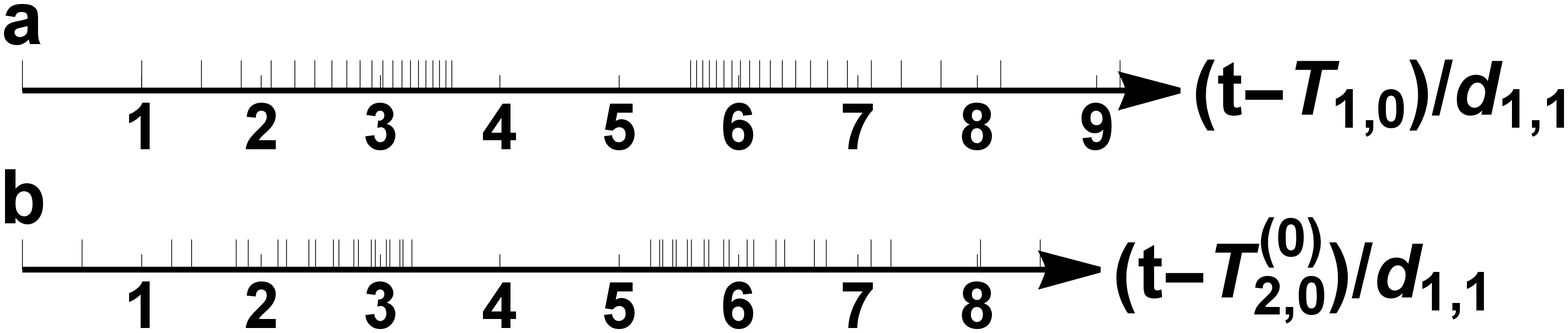}
\caption{Proposal for observing the quantum phase timing of the NCPs between
the (a) first and secondnd and (b) second and third main pulses at $n=20,%
\protect\varepsilon =1/2,$ $T=2d_{1,1}.$}
\label{f5}
\end{figure}

The ultimate requirement for the MZAI is the zeroing of the Doppler phase (%
\ref{123}), i.e. timing must be chosen in such a way that%
\begin{equation}
\phi _{D}=0.  \label{127}
\end{equation}%
Otherwise, the interference term in the excitation probability will be
washed out when averaged over the momenta. In the absence of NCPs $\left(
n=0\right) $, one will satisfy Eq. (\ref{127}) if $T_{1,0}-2T_{2,0}^{\left(
0\right) }+T_{3,0}^{\left( 0\right) }=0$. But then the quantum phase (\ref%
{124a}) is also zeroed. The situation changes in the presence of NCPs. From
Eq. (\ref{96.2}) one can 
%
express pulses timing through the delays $T_{1,0},T$ and delays between NCPs 
$d_{\zeta ,\beta }$ as%
%
\begin{subequations}
\label{128}
\begin{eqnarray}
T_{2,0}^{\left( 0\right) } &=&T_{1,0}+T+\dsum_{m=1}^{2n}\left(
d_{1,m}+d_{2,-m+1}\right) ,  \label{128a} \\
T_{3,0}^{\left( 0\right) } &=&T_{2,0}^{\left( 0\right)
}+T+\dsum_{m=1}^{2n}\left( d_{2,m}+d_{3,-m+1}\right) ,  \label{128b} \\
T_{\zeta ,m}^{\left( 0\right) } &=&T_{\zeta ,0}^{\left( 0\right)
}+\dsum_{m^{\prime }=1}^{m}d_{\zeta ,m^{\prime }},  \label{128c} \\
T_{\zeta ,-m}^{\left( 0\right) } &=&T_{\zeta ,0}^{\left( 0\right)
}-\dsum_{m^{\prime }=1}^{m}d_{\zeta ,-m^{\prime }+1},  \label{128d}
\end{eqnarray}%
where $m>0$. From these equations one can find that the Doppler phase and
the quantum phase are given by 
\end{subequations}
\begin{subequations}
\label{129}
\begin{eqnarray}
\phi _{D} &=&\dfrac{\mathbf{k}}{M}\cdot \mathbf{p}_{i}\dsum_{m=1}^{2n}mf_{m},
\label{129a} \\
\phi _{q} &=&\omega _{k}\dsum_{m=1}^{2n}m^{2}f_{m},  \label{129b}
\end{eqnarray}%
where 
\end{subequations}
\begin{equation}
f_{m}\equiv -d_{1,m}-d_{2,-m+1}+d_{2,m}+d_{3,-m+1}.  \label{130}
\end{equation}%
It is obvious that for $f_{m}\neq 0$, one can arrange the NCPs in such a way
that the Doppler phase will be equal to $0$, and atomic interference will
occur, but the quantum phase will not disappear. We have calculated and
offer the following possibility. Assume that the NCPs in the second and
fourth sets is a mirror image of the NCPs in the first and third sets, 
\begin{subequations}
\label{131}
\begin{eqnarray}
d_{2,-m+1} &=&d_{1,m},  \label{131a} \\
d_{3,-m+1} &=&d_{2,m}.  \label{131b}
\end{eqnarray}%
The Doppler phase will be guaranteed to be zeroed if 
\end{subequations}
\begin{equation}
f_{m}=\dfrac{\left( -1\right) ^{m}}{m}f.  \label{132}
\end{equation}%
If, moreover, it is required that%
\begin{equation}
d_{2,m}=\left[ 1+\left( -1\right) ^{m}\varepsilon \right] d_{1,m},
\label{133}
\end{equation}%
then from Eqs. (\ref{130}-\ref{132}) for intervals between NCPs one gets%
\begin{equation}
d_{1,m}=\dfrac{d_{1,1}}{m}.  \label{134}
\end{equation}%
The timing of NCPs spaced at intervals (\ref{133}, \ref{134}) is shown in
Fig. \ref{f5} for $\varepsilon =\dfrac{1}{2}.$

\begin{figure}[t]
\includegraphics[width=8cm]{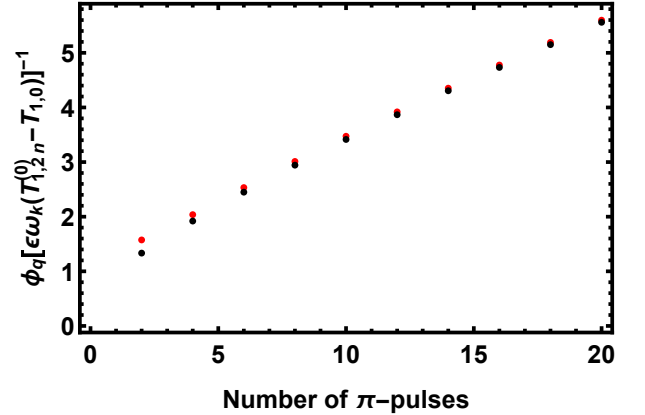}
\caption{Quantum phase as a function of the number of NCPS. Exact (\protect
\ref{135}) and asymptotic (\protect\ref{136}) dependences are plotted in
black and red, respectively.}
\label{f6}
\end{figure}

Here one arrives at the expression for the quantum phase%
\begin{equation}
\phi _{q}=\varepsilon \omega _{k}\left( T_{1,2n}^{\left( 0\right)
}-T_{1,0}\right) \dfrac{2n}{\psi \left( 2n+1\right) +\gamma },  \label{135}
\end{equation}%
where $\psi \left( z\right) $ is the digamma Euler function and $\gamma $ is
the Euler constant. For large $n$, the quantum phase grows as%
\begin{equation}
\phi _{q}=\varepsilon \omega _{k}\left( T_{1,2n}^{\left( 0\right)
}-T_{1,0}\right) \dfrac{2n}{\ln 2n+\gamma }.  \label{136}
\end{equation}%
The quantum phase is shown in Fig. \ref{f6}

\subsection{\label{s3.1}Phase $\bar{\protect\phi}_{q}$}

%
If the recoil frequency is comparable to the inverse pulse duration, then a
smooth quantum phase dependence appears not on the distance between the
pulses, but on the pulse duration. It arises owing to the fact that the
amplitude of the atom in the ground state $\left\vert g\right\rangle $ on
the nonresonant branch of the recoil diagram changes its phase, remaining
unchanged in absolute value. The changes are due to the fact that the
diagonal elements of the $s-$matrices (\ref{71a}, \ref{86a}, \ref{94a})
contain phase factors and also to the fact that the pulse duration depends
on the Raman detuning on the nonresonant branch [see Eq. (\ref{94.1})],
which, according to Eq. (\ref{108}), determined by the recoil frequency. In
the case of rectangular pulses, we calculated these changes [see phase
augmentation in Eqs. (\ref{111b}, \ref{111d}, \ref{111g}, \ref{111i})].%
%

If all NCPs belong to the same class, $\ell _{\zeta ,\beta }=const$, then
from Eqs. (\ref{83c}, \ref{92c}, \ref{110}, \ref{94.1}, \ref{124b}) one
could make sure that the phase $\bar{\phi}_{q}$ does not depend on recoil
frequency , 
\begin{subequations}
\label{137}
\begin{gather}
\left. \bar{\phi}_{q}\right\vert _{\ell _{\zeta ,\beta }=1}=\dfrac{\pi }{2}%
\dsum_{m=1}^{2n}\left( -1\right) ^{m}\left[ -\sqrt{4j_{1,m,1}^{2}-1}\right. 
\notag \\
\left. -\sqrt{4j_{2,-m,1}^{2}-1}+\sqrt{4j_{2,m,1}^{2}-1}+\sqrt{%
4j_{3,-m,1}^{2}-1}\right] ,  \label{137a} \\
\left. \bar{\phi}_{q}\right\vert _{\ell _{\zeta ,\beta }=2}=\pi
\dsum_{m=1}^{2n}\left( -1\right) ^{m}\left( -j_{1,m,2}-j_{2,-m,2}\right. 
\notag \\
\left. +j_{2,m,2}+j_{3,-m,2}\right) ,  \label{137b} \\
\left. \bar{\phi}_{q}\right\vert _{\ell _{\zeta ,\beta }=3_{\pm }}=0.
\label{137c}
\end{gather}%
In these equations, we omitted terms that are multiples of $2\pi $. To
obtain a phase dependent on $\omega _{k}$, it is necessary that at least one
NCP differs from others in the number of rectangular pulses included in it.
We have considered the case when for all NCPs $\ell _{\zeta ,\beta }=2$,
except for the NCP $\left\{ 1.2n\right\} $, for which $\ell _{1.2n}=3_{\pm }$%
. From Eqs. (\ref{82}, \ref{83c}, \ref{91b}, \ref{92c}, \ref{124b}) one
arrives at the result 
\end{subequations}
\begin{subequations}
\label{138}
\begin{gather}
\bar{\phi}_{q}=\left. \bar{\phi}_{q}\right\vert _{\ell _{\zeta ,\beta }=2}+%
\dfrac{\pi }{2}+\Phi _{\pm }\left( 4n\omega _{k}\right) ,  \label{138a} \\
\Phi _{\pm }\left( \nu \right) =  \notag \\
-2\arctan \left[ \dfrac{2\nu \tau }{\sqrt{\pi ^{2}+4\nu ^{2}\tau ^{2}}}\tan
\left( \dfrac{1}{4}\sqrt{\pi ^{2}+4\nu ^{2}\tau ^{2}}\right) \right]  \notag
\\
+\arctan \left( \dfrac{2\nu \tau }{\sqrt{\pi ^{2}+4\nu ^{2}\tau ^{2}}}\tan 
\dfrac{\tau _{1}}{4\tau }\sqrt{\pi ^{2}+4\nu ^{2}\tau ^{2}}\right)  \notag \\
+\arctan \dfrac{\func{Im}\left( s_{d1}s_{d2}\right) }{\func{Re}\left(
s_{d1}s_{d2}\right) }  \notag \\
\mp \left[ \arccos \dfrac{1}{2}\left\vert \dfrac{s_{a2}}{s_{a1}s_{d1}}%
\right\vert +\arccos \dfrac{\left\vert s_{a2}\right\vert \left( \left\vert
s_{a1}\right\vert ^{2}-\left\vert s_{d1}\right\vert ^{2}\right) }{%
2\left\vert s_{a1}s_{d1}s_{d2}\right\vert }\right]  \label{138b}
\end{gather}%
The dependencies (\ref{138b}) are shown in Fig. \ref{f7}.

\begin{figure}[!t]
\includegraphics[width=8cm]{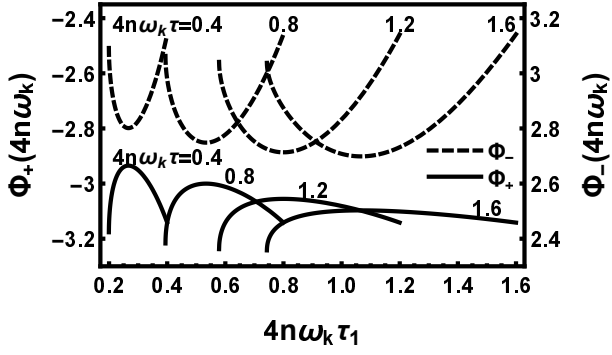}
\caption{Smooth dependence of the phase as a function of the duration of the
first rectangular pulse in the NCP $\protect\tau _{1}$ for different values
of the parameter $4n\protect\omega _{k}\protect\tau .$}
\label{f7}
\end{figure}

\onecolumngrid%

\section{\label{sV}Gravity phase}

%
The main reason for using SLMT is the increase in the gravitational phase of
the AI. For this phase, from Eqs. (\ref{128}, \ref{125a}) one obtains.%
\end{subequations}
\begin{eqnarray}
\phi _{g}^{\left( 0\right) } &=&\left( \mathbf{k\cdot g}-\alpha \right)
\left\{ \left( 2n+1\right) T^{2}+T\dsum_{m=1}^{2n}\left[ \left(
2n+1-m\right) d_{2,-m+1}+\left( 2n+1+m\right) d_{2,m}+2md_{3,-m+1}\right]
+\right.  \notag \\
&&+\dfrac{2n+1}{2}\left[ \left( \dsum_{m=1}^{2n}\left(
d_{2,m}+d_{3,-m+1}\right) \right) ^{2}-\left( \dsum_{m=1}^{2n}\left(
d_{1,m}+d_{2,-m+1}\right) \right) ^{2}\right]  \notag \\
&&+\dfrac{1}{2}\dsum_{m=1}^{2n}\left[ \left( \dsum_{m^{\prime
}=1}^{m}d_{1,m^{\prime }}\right) ^{2}-\left( \dsum_{m^{\prime
}=1}^{m}d_{2,-m^{\prime }+1}\right) ^{2}-\left( \dsum_{m^{\prime
}=1}^{m}d_{2,m^{\prime }}\right) ^{2}+\left( \dsum_{m^{\prime
}=1}^{m}d_{3,-m^{\prime }+1}\right) ^{2}\right]  \notag \\
&&+\left[ \dsum_{m=1}^{2n}\left( d_{1,m}+d_{2,-m+1}\right) \right] \left[
\dsum_{m=1}^{2n}\left( \left( 2n+1-m\right) d_{2,-m+1}+\left( 2n+1+m\right)
d_{2,m}+2md_{3,-m+1}\right) \right]  \notag \\
&&\left. -\left[ \dsum_{m=1}^{2n}\left( 2n+1-m\right) d_{3,-m+1}\right] %
\left[ \dsum_{m^{\prime }=1}^{2n}\left( d_{2,m}+d_{3,-m+1}\right) \right]
\right\} .  \label{139}
\end{eqnarray}

Let us go to the calculation of the correction (\ref{125b}). In the timing
of a given Raman pulse (\ref{96.3}), the $\Upsilon _{\zeta ,\beta }$ part is
the sum of the durations of all preceding pulses, i.e.%
\begin{equation}
\Upsilon _{\zeta ,m}=\left\{ 
\begin{array}{l}
\dsum_{m^{\prime }=0}^{m-1}\tau \left( \nu ,1,m^{\prime },\ell _{1,m^{\prime
}}\right) ,\text{ for }\zeta =1 \\ 
\Upsilon _{\zeta -1,2n}+\tau \left( \nu ,\zeta -1,2n,\ell _{\zeta
-1,2n}\right) +\dsum_{m^{\prime }=-2n}^{m-1}\tau \left( \nu ,\zeta
,m^{\prime },\ell _{\zeta ,m^{\prime }}\right) ,\text{ for }\zeta >1%
\end{array}%
\right.  \label{140}
\end{equation}%
Here it is convenient to pick out the durations of the main pulses (\ref%
{96.1}), so that 
\begin{subequations}
\label{141}
\begin{eqnarray}
\Upsilon _{1,m} &=&\tau +\Theta _{1,m},  \label{141a} \\
\Upsilon _{2,m} &=&\left\{ 
\begin{array}{c}
\tau +\Theta _{2,m},\text{ for }-2n\leq m\leq 0 \\ 
3\tau +\Theta _{2,m},\text{ for }0<m\leq 2n%
\end{array}%
\right. ,  \label{141b} \\
\Upsilon _{3,m} &=&3\tau +\Theta _{3,m}  \label{141c}
\end{eqnarray}%
Calculations bring one to the next result 
\end{subequations}
\begin{subequations}
\label{142}
\begin{eqnarray}
\phi _{g}^{\left( 1\right) } &=&\phi _{g\tau }+\phi _{ga},  \label{142a} \\
\phi _{g\tau } &=&\left( \mathbf{k\cdot g}-\alpha \right) \tau \left\{
T\left( 4n+2+\dfrac{4}{\pi }\right) +2\dsum_{m=1}^{2n}\left[ \dfrac{1}{\pi }%
\left( d_{1,m}+d_{2,-m+1}\right) +\left( \dfrac{1}{\pi }+m\right) \left(
d_{2,m}+d_{3,-m+1}\right) \right] \right\} ,  \label{142b} \\
\phi _{ga} &=&\left( \mathbf{k\cdot g}-\alpha \right) \left\{
-2T_{2,0}^{\left( 0\right) }\Theta _{2,0}+T_{3,0}^{\left( 0\right) }\Theta
_{30}+\dsum_{m=1}^{2n}\left[ \left( \Theta _{1,m}T_{1,m}^{\left( 0\right)
}-\Theta _{2,-m}T_{2,-m}^{\left( 0\right) }-\Theta _{2,m}T_{2,m}^{\left(
0\right) }+\Theta _{3,m}T_{3,-m}^{\left( 0\right) }\right) \right. \right. 
\notag \\
&&+\left( \zeta _{m+1}\tau \left( \left( -1\right) ^{m}2\xi _{m}\omega
_{k},1,m,\ell _{1,m}\right) +\left( -1\right) ^{m}\eta _{1}^{\left( \ell
_{1,m}\right) }\left( \left( -1\right) ^{m}2\xi _{m}\omega _{k}\right)
\right) T_{1,m}^{\left( 0\right) }  \notag \\
&&+\left( \zeta _{m}\tau \left( \left( -1\right) ^{m}2\xi _{m}\omega
_{k},2,-m,\ell _{2,-m}\right) -\left( -1\right) ^{m}\eta _{1}^{\left( \ell
_{2,-m}\right) }\left( \left( -1\right) ^{m}2\xi _{m}\omega _{k}\right)
\right) T_{2,-m}^{\left( 0\right) }  \notag \\
&&+\left( \zeta _{m+1}\tau \left( \left( -1\right) ^{m}2\xi _{m}\omega
_{k},2,m,\ell _{2,m}\right) +\left( -1\right) ^{m}\eta _{1}^{\left( \ell
_{2,m}\right) }\left( 2\left( -1\right) ^{m}\xi _{m}\omega _{k}\right)
\right) T_{2,m}^{\left( 0\right) }  \notag \\
&&\left. \left. +\left( \zeta _{m}\tau \left( \left( -1\right) ^{m}2\xi
_{m}\omega _{k},3,-m,\ell _{3,-m}\right) -\left( -1\right) ^{m}\eta
_{1}^{\left( \ell _{3,-m}\right) }\left( 2\left( -1\right) ^{m}\xi
_{m}\omega _{k}\right) \right) T_{3,-m}^{\left( 0\right) }\right] \right\} .
\label{142c}
\end{eqnarray}%
In the absence of auxiliary NCPs, at $n=0$, one returns to the well-known
result \cite{i31} 
\end{subequations}
\begin{equation}
\phi _{g}=\left( \mathbf{k\cdot g}-\alpha \right) \left[ T^{2}+\tau T\left(
2+\dfrac{4}{\pi }\right) \right] .  \label{143}
\end{equation}%
%
One should note that the correction (\ref{142a}), as well as the
gravitational phase (\ref{139}), grows with the increase in the number of
NCPs.%
%

\twocolumngrid%

\section{ Discussion.}

The model adopted here, a rectangular pulse of the optical field, is widely
used in atomic interferometry and in the theory of atomic clocks \cite{i32}.
At the same time, we are not aware of the consideration of corrections
related to the non-ideal pulse shape. Such corrections arise due to the
non-permanent field amplitude inside the pulse and due to the non-zero
duration of the forward and backward fronts of the pulse $\tau _{f}$. So
instead of one small parameter (\ref{1}) in the Bragg regime, in our case at
least two parameters should be small, 
\begin{subequations}
\label{144}
\begin{eqnarray}
\delta \left\vert \Omega \right\vert /\left\vert \Omega \right\vert &\ll &1,
\label{144a} \\
\omega _{k}\tau _{f} &\ll &1,  \label{144b}
\end{eqnarray}%
where $\delta \left\vert \Omega \right\vert $ is the deviation of the Rabi
frequency from a constant value. The article \cite{i33} reported that the
field intensity was kept constant with an accuracy of 1\%, meaning that 
\end{subequations}
\begin{equation}
\delta \left\vert \Omega \right\vert /\left\vert \Omega \right\vert \sim
5\times 10^{-3}  \label{145}
\end{equation}%
can be implemented.

In Ref. \cite{i34} the front durations were 10ns; with a typical value of $%
\omega _{k}\sim 10^{5}$s$^{-\text{1}}$ one has the estimate%
\begin{equation}
\omega _{k}\tau _{f}\sim 10^{-3}.  \label{146}
\end{equation}%
The fact that the small parameters (\ref{144}) are 40 or 200 times smaller
than the parameter (\ref{1}) allows us to hope that the SLMT option proposed
here is feasible.

It should be noted that the durations of the fronts cannot decrease
indefinitely, since for the applicability of the adiabatic elimination of
the upper level amplitude in Eq. (\ref{28}), the front duration must be long
enough,%
\begin{equation}
\Delta \tau _{f}\gg 1.  \label{147}
\end{equation}%
With a typical value of $\Delta \sim 2\pi \times 1$GHz, $\Delta \tau
_{f}\sim 60$. If, however, one will be able to create pulses with picosecond
fronts, then in order to use them in atomic interferometry, one must first
increase the one-photon detuning $\Delta $ and, accordingly, the field
intensity also needs to be increased.

We predict that the MZAI quantum phase will not disappear in our case. It
arises due to the phase (\ref{45d}) in the Schr\"{o}dinger equation (\ref{44}%
), which is also valid in both Bragg (\ref{1}) and Raman-Nath (\ref{3.1})
regimes. Nevertheless, the quantum phase was not observed in either the
Bragg regime in \cite{i10} or the Raman-Nath regime in \cite{i14}. One can
explain it by the fact that in those articles the auxiliary pulses were
timed out equidistant and in this case the combination of delays between
NCPs in Eq. (\ref{130}) $f_{m}=0$. If one places the NCPs nonequidistant,
then our expression (\ref{129b}) for the quantum phase $\phi _{q}$ can be
directly used in the Bragg regime \cite{i10}. Here, for example, one can use
the nonequidistant NCPs timing shown in Fig. \ref{f5}.

Our result cannot be used in the Raman-Nath regime, since in this case the
atomic momentum in the recoil diagram changes along both branches. An
example of such a diagram is shown in \cite{i14}. In the Raman-Nath regime,
the calculation of the quantum phase must be carried out again. We hope to
carry out this calculation in the future.

If the time budget for auxiliary NCPs is comparable to the interrogation
time $T$, then the quantum phase (\ref{135}) at $n\mathbf{\thicksim }%
\varepsilon \mathbf{\thicksim }1$ is comparable to the maximal expected
quantum phase (\ref{19b}). This is the fundamental difference between our
case and the quantum phases considered in Refs. \cite{i21,i22,i23,i24,i25},
where the quantum phases were only small additions.

Like the gravity phase (\ref{5b}), the quantum phase grows with an increase
in the number of auxiliary NCPs, i.e., with an increase in momentum
transfer. However, since the momentum transfer occurs gradually, increasing
by $\hbar \mathbf{k}$ under the action of each NCP, the quantum phase grows
more slowly than the gravity one in the factor $\ln n$ [see Eq. (\ref{136})].

In this work, as in other papers, we assumed that the interrogation time $T$
is the same between the first and second and between the second and third
main Raman pulses. The SLMT technique allows us to make these times
different, since the resulting Doppler phase can be compensated by Doppler
phases during the operation of auxiliary NCPs. With such an opportunity, a
quantum phase should also arise. We hope to consider this option in the
future.

The quantum phase (\ref{135}) is linear in time. This is not the only phase
linear in time. Another linear phase observed by B. Young \cite{i30} is the
Ramsey phase (\ref{122}). In order to extract the quantum phase, we propose
to scale all the delay times between pulses into a factor $1+x$,%
\begin{equation}
\left\{ T_{1,0},d_{\zeta ,\beta }\right\} \rightarrow \left( 1+x\right)
\left\{ T_{1,0},d_{\zeta ,\beta }\right\} .  \label{148}
\end{equation}%
If simultaneously one scales the Raman detunings as%
\begin{equation}
\delta _{\zeta ,\beta }\rightarrow \delta _{\zeta ,\beta }/\left( 1+x\right)
,  \label{149}
\end{equation}%
then the Ramsey phase (\ref{122}) remains unchanged, only the quantum phase
grows linearly in $x$, and the excitation probability $w$ is a periodic
function of $x$ with period%
\begin{equation}
\Delta x=2\pi \left[ \varepsilon \omega _{k}\left( T_{1,2n}^{\left( 0\right)
}-T_{1,0}\right) \dfrac{2n}{\psi \left( 2n+1\right) +\gamma }\right] ^{-1}.
\label{150}
\end{equation}%
To avoid violating the resonance condition (\ref{100.1}) during scaling, the
parameter $x$ must be small,%
\begin{equation}
x\ll \left( \omega _{k}\tau \right) ^{-1}.  \label{151}
\end{equation}%
Since even for $n\mathbf{\thicksim 1}$ and $\varepsilon \mathbf{\thicksim 1}$%
\begin{equation}
\Delta x\omega _{k}\tau \sim \dfrac{\tau }{T_{1,2n}^{\left( 0\right)
}-T_{1,0}}\ll 1,  \label{152}
\end{equation}%
one can observe many periods of quantum oscillations of the excitation
probability without significantly violating the resonance condition.

We predict a new quantum effect, the dependence of the phase on the pulse
durations $\tau $ and $\tau _{1}$, the term $\bar{\phi}_{q}$. Unlike the
phase linear in $T$ (\ref{19b}), the term $\bar{\phi}_{q}$ is a non-linear
function of $\left\{ \tau ,\tau _{1}\right\} $. Another difference from the
term (\ref{19b}) is that it is specific only to the variant of SLMT
considered here. Neither in the Bragg regime (\ref{1}) nor in the Raman-Nath
regime (\ref{3.1}) does the term $\bar{\phi}_{q}$ occur. Its appearance is
due to the fact that the Raman frequency detuning on the nonresonant branch
of the recoil diagram is proportional to the recoil frequency [see Eq. (\ref%
{108})]. This leads to the fact that in the case of NCPs of type $\ell =1$,
the pulse duration at which, owing to Rabi oscillations, the atom is not
excited on the nonresonant branch, also depends on $\omega _{k}$, see Eqs. (%
\ref{69}, \ref{108}). If the NCPs type is $\ell >1$, then the frequency
detuning (\ref{108}) is the atomic coherence nutation frequency in the space
between pulses. Therefore, the delay between pulses $\tau _{b}$ also depends
on $\omega _{k}$ [see Eqs.(\ref{82}, \ref{91b}, \ref{108})]. Despite that
the atoms remain in the ground state on the nonresonant branch of the recoil
diagram, the phase of the amplitude of this state changes [see Eqs. (\ref{83}%
, \ref{92})] and this change also contributes to the $\bar{\phi}_{q}$ phase.
If the rapidly oscillating quantum phase $\phi _{q}$ vanishes at an
equidistant location of the NCPs, then the smooth phase $\bar{\phi}_{q}$
also ceases to depend on the recoil frequency if all NCPs are of the same
type, see Eqs. (\ref{137}). It is necessary to use NCPs of different types.
In Sec. \ref{s3.1} we carried out the calculation in the case when all NCPs
are of type $\ell =2$, except for NCP $\left\{ 1,2n\right\} $, whose type is 
$\ell _{1,2n}=3$.

Another effect that disappears in the convenient MZAI but appears when
additional beam splitters are turned on is the gravitational redshift \cite%
{i31.2}.

Finally, the gravity phase (\ref{125a}), being quadratic in the Raman
pulses' timing, apart from the main term, the first term in curly brackets
in Eq. (\ref{139}), contains cross terms, combinations of interrogation time
and delays $d_{\zeta ,\beta }$ between auxiliary NCPs, and terms quadratic
in $d_{\zeta ,\beta }$. All these terms are pieced together in Eq. (\ref{139}%
). From a mathematical point of view, the gravity phase is caused by the
phase terms (\ref{45d}) in the Schr\"{o}dinger equation (\ref{44}). Since
this equation is valid for any pulse shape $f\left( t\right) $, our result (%
\ref{139}) will also be correct in the Bragg regime, under the conditions of
the experiment in \cite{i10} .

The correction associated with the finite duration of the Raman pulse, on
the contrary, depends on the pulse shape \cite{i35}. Therefore, the result (%
\ref{142}) is only correct for rectangular NCPs.

In this work, we considered only MZAIs. The SLMT method can lead to a
significant increase in the AI phase in other cases as well, for the
asymmetric MZAI \cite{i19}, for atomic two-loop gyroscopes \cite{i36} and
gravity gradiometers \cite{i37}. Calculations of the SLMT technique for
these AIs are left for future work. As for the two-loop AIs, as shown in
Ref. \cite{i22}, the AI response occurs simultaneously with the stimulated
echo response, the phase of which is sensitive to gravity acceleration. Two
methods have been proposed to resolve this problem, the adjustable momentum
transfer \cite{i22} and the time-skewed pulse sequence \cite{i38}. For
atomic gyroscopes, both methods have been implemented in \cite{i39} and \cite%
{i38,i40,i41}. For the atomic gravity gradiometer, only the time-skewed
method was used. However, even a small distortion in time led to the
appearance of a significant background proportional to the gravity
acceleration \cite{i37}. No background should occur in the adjustable
momentum transfer method.

\acknowledgments

I am grateful to Mark Kasevich for discussing the work at an early stage,
and to Brenton Young for bringing to me the observation of the Ramsey phase
in his MZAI.. I am also grateful to S. Kahn, A. Kumarakrishnan, V. I. Yudin,
and A. V. Taichenashev for fruitful discussions.

\appendix

\section{\label{A1}Higher order density harmonics.}

The atomic resonant Kapitza-Dirac effect \cite{i3'} in the field of a
standing wave and its analogs lie at the heart of many atomic beam
splitters. The momentum transfer to an atom, which is a multiple of $\hbar k$%
, and the subsequent interference of atomic states with different momenta
leads to the appearance of higher harmonics of the atomic density. Density
harmonics with a period up to $\lambda /10$, where $\lambda $ is the
standing wavelength, have been observed in AI \cite{i3.0}. Modifications of
the standing wave, i.e., the triangular potential \cite{i3.0.1} and the
bichromatic standing wave \cite{i3.0.2}, have been proposed; moreover,
transfer of momentum to an atom $\pm 21\mathbf{\hbar }k$ was observed \cite%
{i3.0.1}. Despite the scattering of atoms at large angles, the scattering
indicatrix contains not only the desired states $\pm n\mathbf{\hbar }k$, but
also neighboring states \ldots $\left( \pm n-2\right) \hbar \mathbf{k},$ $%
\left( \pm n-1\right) \hbar \mathbf{k},$ $\left( \pm n+1\right) \hbar 
\mathbf{k},$ $\left( \pm n+2\right) \hbar \mathbf{k}\ldots $ The asymptotes
for $n\mathbf{\gg 1}$ inner tails of this indicatrix were obtained in Refs. 
\cite{i3.0.3, i3.0.4}. A complete indicatrix for $3$ types of optical
potentials was obtained in Ref. \cite{i3.0.5}, where it was also shown that
due to neighboring momentum states, the interference pattern with a period
of $\lambda /2n$ has a smooth envelope with a period of $\lambda /2$, and
because of this, it is obvious that the possibility of using an AI of this
type is doubtful, both for precision measurements and for nanolithography.
To get rid of this difficulty, one can use \cite{i26} the Stern-Gerlach
effect \cite{i3.0.6}, i.e., an atom scattering in a magnetic field having a
uniform gradient. It has been shown that one can obtain an atomic lattice
with a period of $100$ nm and a smooth envelope of size $\mathbf{\thicksim }%
100\mu $, which arises owing to the weak inhomogeneity of the magnetic-field
gradient. Another multicolor scheme was proposed in Ref. \cite{i3.0.7},
where the beam splitter consisted of tree traveling waves with frequencies
and wave vectors $\left\{ \Omega ,\mathbf{k}\right\} ,\left\{ \Omega +\delta
_{1},-\mathbf{k}\right\} ,\left\{ \Omega +\delta _{2},-\mathbf{k}\right\} $.
If the frequency detunings are chosen in such a way that $n_{1}\delta
_{1}+n_{2}\delta _{2}=0$, then this combination of fields creates an
amplitude or phase diffraction grating for atoms with a period $\lambda
/2\left( n_{1}+n_{2}\right) $. If the standing wave is replaced by two
counterpropagating waves in the lin$\perp $lin configuration, then such a
field will be a diffraction phase grating for atoms with a period of $%
\lambda /4$ \cite{i3.0.8,i3.0.9}. Note also that the Raman standing wave
method was proposed \cite{i45}. This technique is now widely known as the
double-diffraction scheme \cite{i46}. The main point of the method is that
two Raman pulses with opposite effective wave vectors $\pm \mathbf{k}$ lead
to splitting of the initial momentum state $\left\vert g,\mathbf{p}%
\right\rangle $ into two states $\left\vert e,\mathbf{p}\pm \hbar \mathbf{k}%
\right\rangle $. Since $k\approx 4\pi /\lambda $, then the interference
between the scattered states leads to density modulation, the atomic lattice
with a period $\lambda /4$. If one of the Raman pulses has the configuration
lin$\parallel $lin and the configuration of the other is lin$\perp $lin,
then the Raman standing wave induces an atomic lattice with a period $%
\lambda /8$ \cite{i45}. Scattering potentials with period $\lambda /8$ were
calculated for various components of the hyperfine splitting of $^{\text{85}%
} $Rb \cite{i47}. One can expect that the Raman standing wave method, in
combination with the multicolor technique, produces density harmonics with a
period $\lambda /8n.$


\begin{thebibliography}{99}
\bibitem{i26.1} The inequality (\ref{26b}) means that during the time $%
\Delta ^{-1}$ one can neglect the changes in the phase of the wave function
associated with the Raman frequency detuning $\tilde{\delta}$, with the
Doppler shift (\ref{18b}) and the change in this shift due to the
acceleration of the atom $\left\vert \mathbf{q}_{i}\cdot \mathbf{g}%
\right\vert T$, with the recoil frequency (\ref{2}), with characteristic
times of change in the amplitude and phase of the fields $\tau _{f}$ and $%
\dot{\phi}^{-1}$. Under these conditions, one can eliminate the first term
on the right-hand side of Eq. (\ref{25c}) and then, using integration by
parts, arrive at the expression (\ref{28}) for the level $\left\vert
o\right\rangle $ amplitude.
\bibitem{i27} Another possibility here is to place detunings $\Delta $
between sublevels of the hyperfine structure of the state $o$, so that the
contributions to the ac-Stark shift from different sublevels cancel each
other.
\end{thebibliography}
\end{document}